\documentclass[journal]{IEEEtran}
\usepackage{makeidx}
\usepackage{amsfonts}
\usepackage{cite}
\usepackage{epsfig,dsfont,amsmath,amssymb}
\usepackage{graphicx,wrapfig,floatflt}
\usepackage{theorem}
\usepackage{url}

\def \func {\mathrm}
\def \bs {\boldsymbol}
\def \Pr {\mathbb{P}}
\def \hn {\mathrm{H}_0}
\def \ha {\mathrm{H}_1}
\def \tr {\mathrm{tr}}

\def \diag {\mathrm{diag}}
\def \rank {\mathrm{rank}}
\def \eye {\mathrm{I}}
\newtheorem{lm}{Lemma}
\newtheorem{thm}{Theorem}
\newtheorem{cor}{Corollary}
\newtheorem{pro}{Proposition}
\begin{document}
\title{Performance Analysis for Sparse Support\\ \vskip -0.35cm Recovery}
\author{Gongguo~Tang,~\IEEEmembership{Student Member,~IEEE,}~ Arye~Nehorai,~%
\IEEEmembership{Fellow,~IEEE}\thanks{This work was supported by
the Department of Defense under the Air Force Office of Scientific
Research MURI Grant FA9550-05-1-0443, and ONR Grant
N000140810849.}\\
\thanks{G. Tang and A. Nehorai are with the Department of Electrical and Systems Engineering, Washington University in St. Louis, St. Louis, MO 63130 USA.}}
 \maketitle

\begin{abstract}
The performance of \mbox{estimating} the common
support for jointly sparse signals based on their projections onto
lower-dimensional space is analyzed. Support recovery is
formulated as a \mbox{multiple-hypothesis} testing problem. Both upper and lower bounds on the probability of error are
derived for general measurement matrices, by using the Chernoff
bound and Fano's inequality, respectively. The upper
bound shows that the performance is determined by a quantity measuring the measurement matrix incoherence, while the lower bound reveals the importance of the total
measurement gain. The
lower bound is applied to derive the minimal number of samples
needed for accurate direction-of-arrival (DOA) estimation for a sparse representation based 
algorithm. When applied to Gaussian
measurement ensembles, these bounds give necessary and sufficient
conditions for a vanishing probability of error for
majority realizations of the measurement matrix. Our results offer
surprising insights into sparse signal recovery. For example, as far as support recovery is
concerned, the well-known bound in Compressive Sensing with the Gaussian measurement matrix is
generally not sufficient unless the noise level is low. Our
study provides an alternative performance measure, one that is
natural and important in practice, for signal recovery in
Compressive Sensing and other application areas exploiting signal sparsity.
\end{abstract}

\markboth{To appear on IEEE Trans. Information Theory} {} 

\begin{IEEEkeywords}
\noindent Chernoff bound, Compressive Sensing,  Fano's inequality, jointly sparse signals, multiple hypothesis testing, probability of error, support recovery
\end{IEEEkeywords}

%
\IEEEpeerreviewmaketitle

\section{Introduction}

%
%
%
%

\IEEEPARstart{S}{upport} recovery for jointly sparse signals concerns
accurately estimating the non-zero component locations shared by a set of
sparse signals based on a limited number of noisy linear observations. More
specifically, suppose that $\{x(t)\in \mathbb{F}^{N},t=1,2,\ldots ,T\},\ \mathbb{F%
}=\mathbb{R}\ \text{or}\ \ \mathbb{C},\ $ is a sequence of jointly sparse
signals (possibly under a sparsity-inducing basis $\Phi $ instead of the
canonical domain) with a common support $S$, which is the index set
indicating the non-vanishing signal coordinates. This model is the same as the joint sparsity \mbox{model 2} (JSM-2) in \cite{Wakin2005joint}. The observation model is
linear:
\begin{equation}
y(t)=Ax(t)+w(t)\ \ \ \ t=1,2,\ldots ,T.  \label{model_intro}
\end{equation}%
In \eqref{model_intro}, $A\in \mathbb{F}^{M\times N}$ is the measurement
matrix, $y(t)\in \mathbb{F}^{M}$ the noisy data vector, and $w(t)\in \mathbb{%
F}^{M}$ an additive noise. In most cases, the sparsity level
$K\triangleq |S|$ and the number of observations $M$ is far less
than $N$, the dimension of the ambient space. This problem arises
naturally in several signal processing
areas such as Compressive Sensing \cite%
{Candes2006Uncertainty,Donoho2006Compressed,Candes2005Decoding,Candes2008IntroCS,Baraniuk2007CompressiveSensing}%
, source localization \cite%
{Willsky2005Source,Model2006Signal,Cevher2008Distributed,Cevher2009Bayesian}%
, sparse approximation and signal denoising \cite{Donoho1998Atomic}.

Compressive Sensing \cite%
{Candes2006Uncertainty,Donoho2006Compressed,Candes2005Decoding}, a recently developed field exploiting the sparsity property of most natural signals, shows great promise to reduce signal sampling rate. In the
classical setting of Compressive Sensing, only one snapshot is considered;
\emph{i.e.}, $T=1$ in \eqref{model_intro}. The goal is to recover a long vector $x:=x(1)$ with a
small fraction of non-zero coordinates from the much shorter observation
vector $y:=y(1)$. Since most natural signals are compressible under some
basis and are well approximated by their $K-$sparse representations \cite%
{Mallat1999Wavelet}, this scheme, if properly justified, will reduce the
necessary sampling rate beyond the limit set by Nyquist and Shannon \cite%
{Baraniuk2007CompressiveSensing,Candes2008IntroCS}. Surprisingly, for exact $K-$sparse signals, if $%
M=O(K\log (\frac{N}{K}))\ll N$ and the measurement matrix is
generated randomly from, for example, a Gaussian distribution, we
can recover $x$ exactly in the noise-free setting by solving a
linear programming task. Besides, various methods have been
designed for the noisy case\cite{Gorodnitsky1997Focuss,Candes2007Dantzig,Tropp2007OMP,Dai2008Subspace,Needell2008Cosamp}%
. Along with these algorithms, rigorous theoretic analysis is
provided to
guarantee their effectiveness in terms of, for example, various $l_{p}$%
-norms of the estimation error for $x$ \cite{Gorodnitsky1997Focuss,Candes2007Dantzig,Tropp2007OMP,Dai2008Subspace,Needell2008Cosamp}. However, these results offer no guarantee that we can recover
the support of a sparse signal correctly.

The accurate recovery of signal support is crucial to Compressive Sensing
both in theory and in practice. Since for signal recovery it is necessary to
have $K\leq M$, signal component values can be computed by solving a
least squares problem once its support is obtained. Therefore, support
recovery is a stronger theoretic criterion than various $l_{p}$-norms. In
practice, the success of Compressive Sensing in a variety of applications
relies on its ability for correct support recovery because the non-zero
component indices usually have significant physical meanings. The support of
temporally or spatially sparse signals reveals the timing or location for
important events such as anomalies. The indices for non-zero coordinates in the Fourier
domain indicate the harmonics existing in a signal\cite%
{Borgnat2008Timefrequency}, which is critical for tasks such as spectrum
sensing for cognitive radios \cite{Tian2007Cognitive}. In compressed DNA
microarrays for bio-sensing, the existence of certain target agents in the tested
solution is reflected by the locations of non-vanishing coordinates, while
the magnitudes are determined by their concentrations\cite%
{Baraniuk2007DNA,Vikalo2007DNA,Parvaresh2008DNA,Vikalo2008DNA}. For
compressive radar imaging, the sparsity constraints are usually imposed on
the discretized time--frequency domain. The distance and velocity of an
object have a direct correspondence to its coordinate in the time-frequency
domain. The magnitude determined by coefficients of reflection is of less
physical significance\cite%
{Baraniuk2007Radar,Herman2008Radar,Herman2008Highradar}. In sparse linear
regression \cite{Larsson2007Regression}, the recovered parameter support
corresponds to the few factors that explain the data. In all these
applications, the support is physically more significant than the component
values.

Our study of sparse support recovery is also motivated by the recent
reformulation of the source localization problem as one of sparse spectrum
estimation. In \cite{Willsky2005Source}, the authors transform the process
of source localization using sensory arrays into the task of estimating the
spectrum of a sparse signal by discretizing the parameter manifold. This
method exhibits super-resolution in the estimation of direction of arrival
(DOA) compared with traditional techniques such as beamforming \cite%
{Johnson1993Array}, Capon\cite{Capon1969Capon}, and MUSIC \cite%
{Schmidt1986MUSIC,Bienvenu1980MUSIC}. Since the basic model employed in \cite%
{Willsky2005Source} applies to several other important problems in signal
processing (see \cite{Nehorai1990MUSIC} and references therein), the
principle is readily applicable to those cases. This idea is later
generalized and extended to other source localization settings in \cite%
{Model2006Signal,Cevher2008Distributed,Cevher2009Bayesian}. For source
localization, the support of the sparse signal reveals the DOA of sources.
Therefore, the recovery algorithm's ability of exact support recovery
is key to the effectiveness of the method. We also note that usually
multiple temporal snapshots are collected, which results in a jointly sparse
signal sets as in \eqref{model_intro}. In addition, since $M$ is the number
of sensors while $T$ is the number of temporal samples, it is far more
expensive to increase $M$ than $T$. The same comments apply to several other
examples in the Compressive Sensing applications discussed in the previous
paragraph, especially the compressed DNA microarrays, spectrum sensing for
cognitive radios, and Compressive Sensing radar imaging.

The signal recovery problem with joint sparsity constraint \cite%
{Duarte2005Distributed,Duarte2006DCS,Fornasier2006,Eldar2007Continuous},
also termed the multiple
measurement vector (MMV) problem{\cite%
{Cotter2005Inverse,Chen2005MMV,Chen2006MMV,Tropp2006Greedy,Tropp2006Convex},
has been considered in a line of previous works. Several algorithms, among
them Simultaneous Orthogonal Matching Pursuit (SOMP) \cite%
{Cotter2005Inverse,Tropp2006Greedy, Duarte2006DCS}; convex relaxation \cite%
{Tropp2006Convex}; $\ell _{1}-$minimization \cite{Chen2005MMV,Chen2006MMV};
and M-FOCUSS \cite{Cotter2005Inverse}, are proposed and analyzed, either
numerically or theoretically. These algorithms are multiple-dimension
extensions of their one-dimension counterparts. Most performance measures of
the algorithms are concerned with bounds on various norms of the difference
between the true signals and their estimates or their closely related
variants. The performance bounds usually involve the mutual coherence
between the measurement matrix $A$ and the basis matrix $\Phi $ under which
the measured signals $x(t)$ have a jointly sparse representation. However,
with joint sparsity constraints, a natural measure of performance would be
the model \eqref{model_intro}'s potential for correctly identifying the true
common support, and hence the algorithm's ability to achieve this potential.
As part of their research, J. Chen and X. Huo derived, in a noiseless
setting, sufficient conditions on the uniqueness of solutions to %
\eqref{model_intro} under $\ell _{0}$ and $\ell _{1}$ minimization. In \cite%
{Cotter2005Inverse}, S. Cotter \emph{et. al.} numerically compared the
probabilities of correctly identifying the common support by basic matching
pursuit, orthogonal matching pursuit, FOCUSS, and regularized FOCUSS in the
multiple-measurement setting with a range of SNRs and different numbers of
snapshots. }

The availability of multiple temporal samples offers serval advantages to the single-sample case. As suggested by the upper bound \eqref{bd_multiple} on the probability of error, increasing the number of temporal samples drives the probability of error to zero exponentially fast as long as certain condition on the inconsistency property of the measurement matrix is satisfied. The probability of error is driven to zero by scaling the SNR according to the signal dimension in \cite{Aeron2009fundamental}, which is not very natural compared with increasing the samples, however. Our results also show that under some conditions increasing temporal samples is usually equivalent to increasing the number of observations for a single snapshot. The later is generally much more expensive in practice. In addition, when there is considerable noise and the columns of the measurement matrix are normalized to one, it is necessary to have multiple temporal samples for accurate support recovery as discussed in Section \ref{sec:lower} and Section \ref{sec:gaussian_nece}.

Our work has several major differences compared to related work \cite{Wainwright2007Bound} and \cite{Aeron2009fundamental}, which also analyze the performance bounds on the probability of error for support recovery using information theoretic tools. The first difference is in the way the problem is modeled: In \cite{Wainwright2007Bound, Aeron2009fundamental}, the sparse signal is deterministic with known smallest absolute value of the non-zero components while we consider a random signal model. This leads to the second difference: We define the probability of error over the signal and noise distributions with the measurement matrix fixed; In \cite{Wainwright2007Bound, Aeron2009fundamental}, the probability of error is taken over the noise, the Gaussian measurement matrix and the signal support. Most of the conclusions in this paper apply to general measurement matrices and we only restrict ourselves to the Gaussian measurement matrix in Section \ref{sec:gaussian}. Therefore, although we use a similar set of theoretical tools, the exact details of applying them are quiet different. In addition, we consider a multiple measurement model while only one temporal sample is available in \cite{Wainwright2007Bound, Aeron2009fundamental}. In particular, to get a vanishing probability of error, Aeron \emph{et.al.} \cite{Aeron2009fundamental} require to scale the SNR according to the signal dimension, which has a similar effect to having multiple temporal measurements in our paper. Although the first two differences make it difficult to compare corresponding results in these two papers, we will make some heuristic comments in Section \ref{sec:gaussian}.

The contribution of our work is threefold. First, we introduce a
hypothesis-testing framework to study the performance for multiple support
recovery. We employ well-known tools in statistics and information theory
such as the Chernoff bound and Fano's inequality to derive both upper and
lower bounds on the probability of error. The upper bound we derive is for
the \emph{optimal} decision rule, in contrast to performance analysis for
specific sub-optimal reconstruction algorithms\cite%
{Gorodnitsky1997Focuss,Candes2007Dantzig,Tropp2007OMP,Dai2008Subspace,Needell2008Cosamp}%
. Hence, the bound can be viewed as a measure of the measurement system's
ability to correctly identify the true support. Our bounds isolate important
quantities that are crucial for system performance. Since our analysis is
based on measurement matrices with as few assumptions as possible, the
results can be used as a guidance in system design. Second, we apply these
performance bounds to other more specific situations and derive necessary
and sufficient conditions in terms of the system parameters to guarantee a
vanishing probability of error. In particular, we study necessary conditions
for accurate source localization by the mechanism proposed in \cite%
{Willsky2005Source}. By restricting our attention to Gaussian measurement
matrices, we derive a result parallel to those for classical Compressive Sensing \cite{Candes2006Uncertainty,Donoho2006Compressed}, namely, the
number of measurements that are sufficient for signal reconstruction. Even
if we adopt the probability of error as the performance criterion, we get
the same bound on $M$ as in \cite{Candes2006Uncertainty,Donoho2006Compressed}%
. However, our result suggests that generally it is impossible to
obtain the true support accurately with only one snapshot when there is considerable noise. We also
obtain a necessary condition showing that the $\log \frac{N}{K}$
term cannot be dropped in Compressive Sensing. Last but not least,
in the course of studying the performance bounds we explore the
eigenvalue structure of a fundamental matrix in support recovery
hypothesis testing for both general measurement matrices and the
Gaussian measurement ensemble. These results are of independent
interest.

The paper is organized as follows. In Section \ref{sec:modelandpre}, we introduce the
mathematical model and briefly review the fundamental ideas in hypothesis
testing. Section \ref{sec:upper} is devoted to the derivation of upper bounds on the
probability of error for general measurement matrices. We first derive an
upper bound on the probability of error for the binary support recovery
problem by employing the well-known Chernoff bound in detection theory \cite{Vantrees2001Detection} and extend it to multiple support recovery. We also
study the effect of noise on system performance. In Section \ref{sec:lower}, an
information theoretic lower bound is given by using the Fano's inequality,
and a necessary condition is shown for the DOA problem considered in \cite%
{Willsky2005Source}. We focus on the Gaussian ensemble in Section \ref{sec:gaussian}.
Necessary and sufficient conditions on system parameters for accurate
support recovery are given and their implications discussed. The paper is
concluded in Section \ref{sec:conclusion}.

\section{Notations, Models, and Preliminaries}

\label{sec:modelandpre}

\subsection{Notations}
\label{sec:notations}

We first introduce some notations used throughout this paper. Suppose $x\in
\mathbb{F}^{N}$ is a column vector. We denote by $S=\func{supp}(x)\subseteq
\{1,\ldots ,N\}$ the support of $x$, which is defined as the set of indices
corresponding to the non-zero components of $x$. For a matrix $X$, $S=\func{%
supp}\left( X\right) $ denotes the index set of non-zero rows of $X$. Here
the underlying field $\mathbb{F}$ can be assumed as $\mathbb{R}$ or $\mathbb{%
C}$. We consider both real and complex cases simultaneously. For this
purpose, we denote a constant $\kappa =1/2$ or $1$ for the real or complex
case, respectively.

Suppose $S$ is an index set. We denote by $|S|$ the number of elements in $S$%
. For any column vector $x\in \mathbb{F}^{N}$, $x^{S}\in \mathbb{F}^{N}$ is
the vector in $\mathbb{F}^{|S|}$ formed by the components of $x$ indicated
by the index set $S$; for any matrix $B$, $B^{S}$ denotes the submatrix
formed by picking the rows of $B$ corresponding to indices in $S$, while $%
B_{S}$ is the submatrix with columns from $B$ indicated by $S$. If $I$ and $%
J $ are two index sets, then $B_{J}^{I}=(B^{I})_{J}$, the submatrix of $B$
with rows indicated by $I$ and columns indicated by $J$.

Transpose of a vector or matrix is denoted by $^{\prime }$ while conjugate
transpose by $^{\dagger }$. $A\otimes B$ represents the Kronecker product of
two matrices. For a vector $v$, $\diag(v)$ is the diagonal matrix with the
elements of $v$ in the diagonal. The identity matrix of dimension $M$ is $%
\eye_{M}$. The trace of matrix $A$ is given by $\tr(A)$, the determinant by $%
|A|$, and the rank by $\rank(A)$. Though the notation for determinant is
inconsistent with that for cardinality of an index set, the exact meaning
can always be understood from the context.

Bold symbols are reserved for \emph{random} vectors and matrices. We use $%
\Pr $ to denote the probability of an event and $\mathbb{E}$ the
expectation. The underlying probability space can be inferred from the
context. Gaussian distribution for a random vector in field $\mathbb{F}$
with mean $\mu $ and covariance matrix $\Sigma $ is represented by $\mathbb{F%
}\mathcal{N}\left( \mu ,\Sigma \right) $ . Matrix variate Gaussian
distribution \cite{Gupta1999Matrix} for $\bs Y \in \mathbb{F}^{M \times T}$
with mean $\Theta\in \mathbb{F}^{M\times T}$ and covariance matrix $\Sigma
\otimes \Psi$, where $\Sigma \in \mathbb{F}^{M\times M}$ and $\Psi \in
\mathbb{F}^{T\times T}$, is denoted by $\mathbb{F}\mathcal{N}_{M,T}(\Theta,
\Sigma \otimes \Psi)$

Suppose $\{f_{n}\}_{n=1}^{\infty },\{g_{n}\}_{n=1}^{\infty }$ are two
positive sequences, $f_{n}=o(g_{n})$ means that $\lim_{n\rightarrow \infty }%
\frac{f_{n}}{g_{n}}=0$. An alternative notation in this case is $g_{n}\gg
f_{n}$. We use $f_{n}=O(g_{n})$ to denote that there exists an $N\in \mathbb{%
N}$ and $C>0$ independent of $N$ such that $f_{n}\leq Cg_{n}$ for $n\geq N$.
Similarly, $f_{n}=\Omega (g_{n})$ means $f_{n}\geq Cg_{n}$ for $n\geq N$.
These simple but expedient notations introduced by G. H. Hardy greatly
simplify derivations \cite{Chow1997Probability}.

\subsection{Models}
\label{sec:model}
Next, we introduce our mathematical model. Suppose $\boldsymbol{x}\left(
t\right) \in \mathbb{F}^{N},t=1,\ldots ,T$ are jointly sparse signals with
common support; that is, only a few components of $\boldsymbol{x}\left(
t\right) $ are non-zero and the indices corresponding to these non-zero
components are the same for all $t=1,\ldots ,T$. The common support $S=\func{%
supp}\left( \boldsymbol{x}\left( t\right) \right) $ has known size $K=|S|$.
We assume that the vectors $\boldsymbol{x}^{S}\left( t\right) ,t=1,\ldots ,T$
formed by the non-zero components of $\bs x(t)$ follow \textit{i.i.d.} $%
\mathbb{F}\mathcal{N}(0,\eye_{K})$. The measurement model is as follows:%
\begin{equation}
\boldsymbol{y}\left( t\right) =A\boldsymbol{x}\left( t\right) +\boldsymbol{w}%
\left( t\right) ,t=1,2,\ldots ,T,  \label{model_scalar}
\end{equation}%
where $A$ is the measurement matrix and $\boldsymbol{y}\left( t\right) \in
\mathbb{F}^{M}$ the measurements. The additive noise $\boldsymbol{w}\left(
t\right) \in \mathbb{F}^{N}$ is assumed to follow \textit{i.i.d.} $\mathbb{F%
}\mathcal{N}\left( 0,\sigma ^{2}\eye_{M}\right) $. Note that assuming unit
variance for signals loses no generality since only the ratio of signal
variance to noise variance appears in all subsequence analyses. In this
sense, we view $1/\sigma ^{2}$ as the signal-to-noise ratio (SNR).

Let $\boldsymbol{X=}%
\begin{bmatrix}
\boldsymbol{x}\left( 1\right) & \boldsymbol{x}\left( 2\right) & \cdots &
\boldsymbol{x}\left( T\right)%
\end{bmatrix}%
$ and $\boldsymbol{Y}$, $\boldsymbol{W}$ be defined in a similar manner.
Then we write the model in the more compact matrix form:
\begin{equation}
\boldsymbol{Y}=A\boldsymbol{X}+\boldsymbol{W}.  \label{model_vector}
\end{equation}%
We start our analysis for general measurement matrix $A$. For an
arbitrary measurement matrix $A\in \mathbb{F}^{M\times N}$, if
every $M\times M$ submatrix of $A$ is non-singular, we then call
$A$ a \mbox{\emph{non-degenerate}} measurement matrix. In this
case, the corresponding linear system $A x = b$ is said to have
the \emph{Unique Representation Property (URP)}, the implication
of which is discussed in \cite{Gorodnitsky1997Focuss}. While most
of our results apply to general non-degenerate measurement
matrices, we need to impose more structure on the measurement
matrices in order to obtain more profound results. In particular,
we will consider Gaussian measurement matrix $\boldsymbol{A}$
whose elements $\boldsymbol{A}_{mn}$ are generated from \textit{i.i.d.} $%
\mathbb{F}\mathcal{N}(0,1)$. However, since our performance analysis is
carried out by conditioning on a particular realization of $\boldsymbol{A}$,
we still use non-bold $A$ except in Section \ref{sec:gaussian}. The role played
by the variance of $\boldsymbol{A}_{mn}$ is indistinguishable from that of a
signal variance and hence can be combined to $1/\sigma ^{2}$, the SNR, by
the note in the previous paragraph.

We now consider two hypothesis-testing problems. The first one is a binary
support recovery problem:
\begin{equation}
\left\{
\begin{array}{c}
\mathrm{H}_{0}:\func{supp}\left( \boldsymbol{X}\right) =S_{0} \\
\mathrm{H}_{1}:\func{supp}\left( \boldsymbol{X}\right) =S_{1}%
\end{array}%
.\right.  \label{BHT_support}
\end{equation}%
The results we obtain for binary binary support recovery
\eqref{BHT_support} offer insight into our second problem: the
multiple support recovery. In the multiple support recovery
problem we choose one among $\binom{N}{K}$ distinct candidate
supports of $\boldsymbol{X}$, which is a multiple-hypothesis
testing problem:
\begin{equation}
\left\{
\begin{array}{ll}
\mathrm{H}_{0}: & \func{supp}\left( \boldsymbol{X}\right) =S_{0} \\
\mathrm{H}_{1}: & \func{supp}\left( \boldsymbol{X}\right) =S_{1} \\
& \ \ \ \ \ \ \vdots \\
\mathrm{H}_{L-1}: & \func{supp}\left( \boldsymbol{X}\right) =S_{L-1}%
\end{array}%
.\right.  \label{MHT_support}
\end{equation}

\subsection{Preliminaries for Hypothesis Testing}
\label{sec:pre}
We now briefly introduce the fundamentals of hypothesis testing. The
following discussion is based mainly on \cite{Vantrees2001Detection}. In a
simple binary hypothesis test, the goal is to determine which of two
candidate distributions is the true one that generates the data matrix (or
vector) $\bs Y$:
\begin{equation}
\left\{
\begin{array}{c}
\mathrm{H}_{0}:\bs Y\sim p\ (\bs Y|\hn) \\
\mathrm{H}_{1}:\bs Y\sim p\ (\bs Y|\ha)%
\end{array}%
.\right.  \label{BHT}
\end{equation}

There are two types of errors when one makes a choice based on the observed
data $\bs Y$. A \emph{false alarm} corresponds to choosing $\mathrm{H}_{1}$
when $\mathrm{H}_{0}$ is true, while a \emph{miss} happens by choosing $%
\mathrm{H}_{0}$ when $\mathrm{H}_{1}$ is true. The probabilities of these
two types of errors are called the probability of a false alarm and the
probability of a miss, which are denoted by
\begin{eqnarray}
P_{\mathrm{F}} &=&\Pr \ (\text{Choose}\ \mathrm{H}_{1}|\hn), \\
P_{\mathrm{M}} &=&\Pr \ (\text{Choose}\ \mathrm{H}_{0}|\ha),
\end{eqnarray}%
respectively. Depending on whether one knows the prior probabilities $\Pr (%
\mathrm{H}_{0})$ and $\Pr (\mathrm{H}_{1})$ and assigns losses to errors,
different criteria can be employed to derive the optimal decision rule. In
this paper we adopt the probability of error with equal prior probabilities
of $\mathrm{H}_{0}$ and $\mathrm{H}_{1}$ as the decision criterion; that is,
we try to find the optimal decision rule by minimizing
\begin{equation}
P_{\mathrm{err}}=P_{\mathrm{F}}\Pr (\mathrm{H}_{0})+P_{\mathrm{M}}\Pr (\mathrm{H}_{1})=\frac{1%
}{2}P_{\mathrm{F}}+\frac{1}{2}P_{D}.
\end{equation}%
The optimal decision rule is then given by the \emph{likelihood ratio test}:
\begin{equation}
\ell (\bs Y)=\log \frac{p\left( \boldsymbol{Y}|\mathrm{H}_{1}\right) }{%
p\left( \boldsymbol{Y}|\mathrm{H}_{0}\right) }\overset{\mathrm{H}_{1}}{%
\underset{\mathrm{H}_{0}}{\gtrless }}0  \label{LRT}
\end{equation}
where $\log (\cdot)$ is the natural logarithm function.

The probability of error associated with the optimal decision rule, namely, the
likelihood ratio test \eqref{LRT}, is a measure of the best performance a system can achieve.
In many cases of interest, the simple
binary hypothesis testing problem \eqref{BHT} is derived from a
signal-generation system. For example, in a digital communication system,
hypotheses $\hn$ and $\ha$ correspond to the transmitter sending digit $0$
and $1$, respectively, and the distributions of the observed data under the
hypotheses are determined by the modulation method of the system. Therefore,
the minimal probability of error achieved by the likelihood ratio test is a
measure of the performance of the modulation method. For the problem
addressed in this paper, the minimal probability of error reflects the
measurement matrix's ability to distinguish different signal supports.

The Chernoff bound\cite{Vantrees2001Detection} is a well-known tight upper
bound on the probability of error. In many cases, the optimum test can be
derived and implemented efficiently but an exact performance calculation is impossible. Even if such
an expression can be derived, it is too complicated to be of practical use.
For this reason, sometimes a simple bound turns out to be more useful in
many problems of practical importance. The Chernoff bound, based on the
moment generating function of the test statistic $\ell (\bs Y)$ \eqref{LRT}, provides an easy way to compute such a bound.

Define $\mu (s)$ as the logarithm of the moment generating function of $\ell
(Y)$:
\begin{eqnarray}
\mu (s) &\triangleq &\log \int_{-\infty }^{\infty }e^{s\ell (\bs Y)}p(\bs Y|%
\hn)d\bs Y  \notag  \label{log_mgf} \\
&=&\log \int_{-\infty }^{\infty }[p(\bs Y|\ha)]^{s}[p(\bs Y|\hn)]^{1-s}d\bs Y.
\end{eqnarray}%
Then the Chernoff bound states that
\begin{eqnarray}
P_{\mathrm{F}} &\leq &\exp [\mu (s_{m})]\leq \exp [\mu (s)],  \label{Chernoff_F} \\
P_{\mathrm{M}} &\leq &\exp [\mu (s_{m})]\leq \exp [\mu (s)],  \label{Chernoff_M}
\end{eqnarray}%
and
\begin{equation}\label{Chernoff_err}
P_{\mathrm{err}}\leq \frac{1}{2}\exp [\mu (s_{m})]\leq
\frac{1}{2}\exp [\mu (s)],
\end{equation}
where $0\leq s\leq 1$ and $s_{m}=\mathrm{argmin}_{0\leq s\leq 1}\mu (s)$. Note that a refined argument gives the constant $1/2$ in \eqref{Chernoff_err} instead of $1$ as obtained by direct application of \eqref{Chernoff_F} and \eqref{Chernoff_M} \cite{Vantrees2001Detection}. We
use these bounds to study the performance of the support recovery problem.

We next extend to multiple-hypothesis testing the key elements of the binary
hypothesis testing. The goal in a simple multiple-hypothesis testing problem
is to make a choice among $L$ distributions based on the observations:
\begin{equation}
\left\{
\begin{array}{ll}
\mathrm{H}_{0}: & \bs Y\sim p\ (\bs Y|\mathrm{H}_{0}) \\
\mathrm{H}_{1}: & \bs Y\sim p\ (\bs Y|\mathrm{H}_{1}) \\
& \vdots \\
\mathrm{H}_{L-1}: & \bs Y\sim p\ (\bs Y|\mathrm{H}_{L-1})%
\end{array}%
.\right.  \label{MHT}
\end{equation}%
Using the total probability of error as a decision criterion and assuming
equal prior probabilities for all hypotheses, we obtain the optimal decision
rule given by
\begin{equation}
\mathrm{H}^{\ast }=\mathrm{argmax}_{0\leq i\leq L-1}p\ (\bs Y|\mathrm{H}%
_{i}).  \label{opt_MHT}
\end{equation}%
Application of the union bound and the Chernoff bound \eqref{Chernoff_err} shows that the total probability of error is bounded as follows:
\begin{eqnarray}
P_{\mathrm{err}} &=&\sum_{i=0}^{L-1}\Pr (\mathrm{H}^{\ast }\neq \mathrm{H}%
_{i}|\mathrm{H}_{i})\Pr (\mathrm{H}_{i})  \notag  \label{Perr} \\
&\leq &\frac{1}{2L}\sum_{i=0}^{L-1}\sum_{\substack{ j=0  \\ j\neq i}}%
^{L-1}\exp [\mu (s;\mathrm{H}_{i},\mathrm{H}_{j})], 0\leq s\leq 1, \label{bound_m}
\end{eqnarray}%
where $\exp [\mu (s;\mathrm{H}_{i},\mathrm{H}_{j})]$ is the moment-generating function in the binary hypothesis testing problem for $\mathrm{H}_{i}$ and $\mathrm{H%
}_{j}$. Hence, we obtain an upper bound
for multiple-hypothesis testing from that for binary hypothesis testing.

\section{Upper Bound on Probability of Error for Non-degenerate Measurement
Matrices}

\label{sec:upper}

In this section, we apply the general theory for hypothesis testing, the
Chernoff bound on the probability of error in particular, to the support
recovery problems \eqref{BHT_support} and \eqref{MHT_support}. We first
study binary support recovery, which lays the foundation for the general
support recovery problem.

\subsection{Binary Support Recovery}
\label{sec:binaryupper}
Under model \eqref{model_vector} and the assumptions pertaining to it,
observations $\bs Y$ follow a matrix variate Gaussian distribution \cite%
{Gupta1999Matrix} when the true support is $S$:
\begin{equation}
\boldsymbol{Y}|S\sim \mathbb{F}\mathcal{N}_{M,T}(0,\Sigma _{S}\otimes \eye%
_{T})
\end{equation}%
with the probability density function (pdf) given by
\begin{equation}
p\ (\boldsymbol{Y}|S)=\frac{1}{(\pi /\kappa )^{\kappa MT}|\Sigma
_{S}|^{\kappa T}}\exp \left[ -\kappa \tr\left( \boldsymbol{Y}^{\dagger
}\Sigma _{S}^{-1}\boldsymbol{Y}\right) \right] ,
\end{equation}%
where $\Sigma _{S}=A_{S}A_{S}^{\dagger }+\sigma ^{2}\eye_{M}$ is the common
covariance matrix for each column of $\bs Y$. The binary support recovery
problem \eqref{BHT_support} is equivalent to a linear Gaussian binary
hypothesis testing problem:
\begin{equation}
\left\{
\begin{array}{c}
\mathrm{H}_{0}:\bs Y\sim \mathbb{F}\mathcal{N}_{M,T}(0,\Sigma
_{S_{0}}\otimes \eye_{T}) \\
\mathrm{H}_{1}:\bs Y\sim \mathbb{F}\mathcal{N}_{M,T}(0,\Sigma
_{S_{1}}\otimes \eye_{T})%
\end{array}%
.\right.
\end{equation}%
From now on, for notation simplicity we will denote $\Sigma _{S_{i}}$ by $%
\Sigma _{i}$. The optimal decision rule with minimal probability of error
given by the likelihood ratio test $\ell (\bs Y)$ \eqref{LRT} reduces to
\begin{equation}\label{lrt_support}
-\kappa \tr\left[ \boldsymbol{Y}^{\dagger }\left( \Sigma
_{1}^{-1}-\Sigma _{0}^{-1}\right) \boldsymbol{Y}\right] -\kappa
T\log \frac{\left\vert \Sigma
_{1}\right\vert }{\left\vert \Sigma _{0}\right\vert }\overset{\mathrm{H}_{1}}%
{\underset{\mathrm{H}_{0}}{\gtreqless }}0.
\end{equation}

To analyze the performance of the likelihood ratio test
\eqref{lrt_support}, we first compute the log-moment-generating
function of $\ell (\bs Y)$ according to \eqref{log_mgf}:
\begin{eqnarray}
&&\mu (s) \\
&=&\log \int \left[ p\left( \boldsymbol{Y}|\mathrm{H}_{1}\right) \right] ^{s}%
\left[ p\left( \boldsymbol{Y}|\mathrm{H}_{0}\right) \right] ^{1-s}d%
\boldsymbol{Y}  \notag \\
&=&\log \Bigg[\frac{1}{(\pi /\kappa )^{\kappa MT}|\Sigma _{1}|^{\kappa
sT}|\Sigma _{0}|^{\kappa \left( 1-s\right) T}}  \notag \\
&&\times \int \exp \left\{ -\kappa \tr\left[ \boldsymbol{Y}^{\dagger }\left(
s\Sigma _{1}^{-1}+\left( 1-s\right) \Sigma _{0}^{-1}\right) \boldsymbol{Y}%
\right] \right\} d\boldsymbol{Y}\Bigg]  \notag \\
&=&\log \Bigg[\frac{\left\vert s\Sigma _{1}^{-1}+\left( 1-s\right) \Sigma
_{0}^{-1}\right\vert ^{-\kappa T}}{|\Sigma _{1}|^{\kappa sT}|\Sigma
_{0}|^{\kappa \left( 1-s\right) T}}\Bigg]  \notag \\
&=&{-\kappa T}\log \left\vert sH^{1-s}+\left( 1-s\right) H^{-s}\right\vert ,\
\ \ \ 0\leq s\leq 1,
\end{eqnarray}%
where $H=\boldsymbol{\Sigma }_{0}^{1/2}\boldsymbol{\Sigma }_{1}^{-1}%
\boldsymbol{\Sigma }_{0}^{1/2}$. The computation of the exact minimizer $%
s_{m}=\mathrm{argmin}_{0\leq s\leq 1}\mu (s)$ is non-trivial and will lead
to an expression of $\mu (s_{m})$ too complicated to handle. When $%
|S_{0}|=|S_{1}|$ and the columns of $A$ are not highly correlated, for
example in the case of $A$ with \emph{i.i.d.} elements, $s_{m}\approx \frac{1%
}{2}$. We then take $s=\frac{1}{2}$ in the Chernoff bounds \eqref{Chernoff_F}%
, \eqref{Chernoff_M}, and \eqref{Chernoff_err}. Whereas the bounds obtained in this way may not be the absolute best ones, they are still valid.

As positive definite Hermitian matrices, $H$ and $H^{-1}$ can be
simultaneously diagonalized by a unitary transformation. Suppose that the
eigenvalues of $H$ are $\lambda _{1}\geq \cdots \geq \lambda
_{k_{0}}>1=\cdots =1>\sigma _{1}\geq \cdots \geq \sigma _{k_{1}}$ and $D=%
\mathrm{diag}[\lambda _{1},\ldots ,\lambda _{k_{0}},1,\ldots ,1,\sigma
_{1},\ldots ,\sigma _{k_{1}}]$. Then it is easy to show that
\begin{eqnarray}
\mu (1/2) &=&{-\kappa T}\log \left\vert \frac{D^{1/2}+D^{-1/2}}{2}\right\vert
\notag  \label{bd_muhalf} \\
&=&-\kappa T\Bigg[\sum_{j=1}^{k_{0}}\log \left( \frac{\sqrt{\lambda _{j}}+{1}/%
\sqrt{\lambda _{j}}}{2}\right)  \notag \\
&&\ \ \ \ \ \ \ \ \ \ +\sum_{j=1}^{k_{1}}\log \left( \frac{\sqrt{\sigma _{j}}+%
{1}/\sqrt{\sigma _{j}}}{2}\right) \Bigg].
\end{eqnarray}

Therefore, it is necessary to count the numbers of eigenvalues of $H$ that
are greater than 1, equal to 1 and less than 1, \emph{i.e.}, the values of $%
k_{0}$ and $k_{1}$ for general non-degenerate measurement matrix $A$. We
have the following theorem on the eigenvalue structure of $H$:

\begin{pro}
\label{eig_count} For any non-degenerate measurement matrix $A,$ let $%
H=\Sigma _{0}^{1/2}\Sigma _{1}^{-1}\Sigma _{0}^{1/2}$, $k_{\mathrm{i}}=|S_{0}\cap
S_{1}|,k_{0}=\left\vert S_{0}\backslash S_{1}\right\vert =\left\vert
S_{0}\right\vert -k_{\mathrm{i}},k_{1}=\left\vert S_{1}\backslash S_{0}\right\vert
=\left\vert S_{1}\right\vert -k_{\mathrm{i}}$ and assume $M\geqslant k_{0}+k_{1}$;
then $k_{0}$ eigenvalues of matrix $H$ are greater than $1$, $k_{1}$ less
than $1$, and $M-\left( k_{0}+k_{1}\right) $ equal to $1$.
\end{pro}

\textbf{Proof:} See Appendix A.

\bigskip For binary support recovery \eqref{BHT_support} with $%
|S_{0}|=|S_{1}|=K$, we have $k_{0}=k_{1}\triangleq k_{\mathrm{d}}$. The subscripts $\mathrm{i}$ and $\mathrm{d}$ in $k_{\mathrm{i}}$ and $k_{\mathrm{d}}$ are short for ``intersection" and ``difference", respectively.
Employing the Chernoff bounds \eqref{Chernoff_err} and Proposition
\ref{eig_count}, we have

\begin{pro}
\label{thm_binarybound} If $M \geq 2k_{\mathrm d}$, the probability of error for the binary support
recovery problem \eqref{BHT_support} is bounded by
\begin{eqnarray}
P_{\mathrm{err}} &\leq &\frac{1}{2}\left[ \frac{\bar{\lambda}_{S_{0},S_{1}}%
\bar{\lambda}_{S_{1},S_{0}}}{16}\right] ^{-\kappa k_{\mathrm{d}}T/2} ,
\label{bd_binary}
\end{eqnarray}%
where $\bar{\lambda}_{S_{i},S_{j}}$ is the geometric mean of the eigenvalues
of $H=\boldsymbol{\Sigma }_{i}^{1/2}\boldsymbol{\Sigma }_{j}^{-1}\boldsymbol{%
\Sigma }_{i}^{1/2}$ that are greater than one.
\end{pro}

\textbf{Proof:} According to \eqref{Chernoff_err} and \eqref{bd_muhalf}, we
have
\begin{eqnarray*}
P_{\mathrm{err}} &\leq &\frac{1}{2}\exp \left[ \mu \left( \frac{1}{2}\right) %
\right] \\
&\leq &\frac{1}{2}\left[ \prod_{j=1}^{k_{\mathrm{d}}}\left( \frac{\sqrt{\lambda _{j}}%
}{2}\right) \prod_{j=1}^{k_{\mathrm{d}}}\left( \frac{1/\sqrt{\sigma _{j}}}{2}\right) %
\right] ^{-\kappa T} \\
&=&\frac{1}{2}\left[ \frac{\left( \prod_{j=1}^{k_{\mathrm{d}}}\lambda _{j}\right)
^{1/k_{\mathrm{d}}}\left( \prod_{j=1}^{k_{\mathrm{d}}}\frac{1}{\sigma _{j}}\right) ^{1/k_{\mathrm{d}}}}{%
16}\right] ^{-\kappa k_{\mathrm{d}}T/2}.
\end{eqnarray*}%
Define $\bar{\lambda}_{S_{i},S_{j}}$ as the geometric mean of the
eigenvalues of $H=\boldsymbol{\Sigma }_{i}^{1/2}\boldsymbol{\Sigma }_{j}^{-1}%
\boldsymbol{\Sigma }_{i}^{1/2}$ that are greater than one. Then obviously we
have $\bar{\lambda}_{S_{0},S_{1}}=\left( \prod_{j=1}^{k_{\mathrm{d}}}\lambda
_{j}\right) ^{1/k_{\mathrm{d}}}$. Since $H^{-1}$ and $\Sigma _{1}^{1/2}\Sigma
_{0}^{-1}\Sigma _{1}^{1/2}$ have the same set of eigenvalues, $1/\sigma
_{j},j=1,\ldots ,k_{\mathrm{d}}$ are the eigenvalues of $\Sigma _{1}^{1/2}\Sigma
_{0}^{-1}\Sigma _{1}^{1/2}$ that are greater than 1. We conclude that $\bar{%
\lambda}_{S_{1},S_{0}}=\left( \prod_{j=1}^{k_{\mathrm{d}}}1/\sigma _{j}\right)
^{1/k_{\mathrm{d}}}$. $\blacksquare $

Note that $\bar{\lambda}_{S_{0},S_{1}}$ and $\bar{\lambda}_{S_{1},S_{0}}$
completely determine the measurement system \eqref{model_vector}'s
performance in differentiating two different signal supports. It must be
larger than the constant 16 for a vanishing bound when more temporal samples
are taken. Once the threshold 16 is exceeded, taking more samples will drive
the probability of error to 0 exponentially fast. From numerical simulations
and our results on the Gaussian measurement matrix, $\bar{\lambda}%
_{S_{i},S_{j}}$ does not vary much when $S_i, S_j$ and $k_{\mathrm{d}}$ change, as long as the
elements in the measurement matrix $A$ are highly uncorrelated. \footnote{Unfortunately, this is not the case when the columns of $A$ are samples from uniform linear sensor array manifold.} Therefore,
quite appealing to intuition, the larger the size $k_{\mathrm{d}}$ of the difference
set between the two candidate supports, the smaller the probability of error.

\subsection{Multiple Support Recovery}\label{sec:multipleupper}

Now we are ready to use the union bound \eqref{bound_m} to study the
probability of error for the multiple support recovery problem %
\eqref{MHT_support}. We assume each candidate support $S_{i}$ has known
cardinality $K$, and we have $L={\binom{N}{K}}$ such supports. Our general
approach is also applicable to cases for which we have some prior
information on the structure of the signal's sparsity pattern, for example
the setup in model-based Compressive Sensing\cite{Baraniuk2008Model}. In
these cases, we usually have $L\ll {\binom{N}{K}}$ supports, and a careful
examination on the intersection pattern of these supports will give a better
bound. However, in this paper we will not address this problem and will
instead focus on the full support recovery problem with $L={\binom{N}{K}}$.
Defining $\bar{\lambda}=\min_{i\neq j}\{\bar{\lambda}_{S_{i},S_{j}}\}$, we
have the following theorem:

\begin{thm}\label{thm_mul_bd}
If $M \geq 2K$ and $\bar{\lambda}>4\left[ K\left( N-K\right) \right] ^{\frac{1}{\kappa T}}$,
then the probability of error for the full support recovery problem %
\eqref{MHT_support} with $|S_{i}|=K$ and $L={\binom{N}{K}}$ is bounded by
\begin{equation}
P_{\mathrm{err}}\leq \frac{1}{2}\frac{\frac{K\left( N-K\right) }{\left( \bar{%
\lambda}/4\right) ^{\kappa T}}}{1-\frac{K\left( N-K\right) }{\left( \bar{%
\lambda}/4\right) ^{\kappa T}}}.  \label{bd_multiple}
\end{equation}
\end{thm}

\textbf{Proof:} Combining the bound in Proposition
\ref{thm_binarybound} and Equation \eqref{bound_m}, we have
\begin{eqnarray*}
P_{\mathrm{err}} &\leq &\frac{1}{2L}\sum_{i=0}^{L-1}\sum_{\substack{ j=1  \\ %
j\neq i}}^{L-1}\left[ \frac{\bar{\lambda}_{S_{i},S_{j}}\bar{\lambda}%
_{S_{j},S_{i}}}{16}\right] ^{-\kappa k_{\mathrm{d}}T/2} \\
&\leq &\frac{1}{2L}\sum_{i=0}^{L-1}\sum_{\substack{ j=1  \\ j\neq i}}%
^{L-1}\left( \frac{\bar{\lambda}}{4}\right) ^{-\kappa k_{\mathrm{d}}T}.
\end{eqnarray*}%
Here $k_{\mathrm{d}}$ depends on the supports $S_{i}$ and $S_{j}$. For fixed $S_{i}$,
the number of supports that have a difference set with $S_{i}$ with
cardinality $k_{\mathrm{d}}$ is ${\binom{K}{k_{\mathrm{d}}}}{\binom{{N-K}}{k_{\mathrm{d}}}}$.
Therefore, using ${\binom{K}{k_{\mathrm{d}}}}\leq K^{k_{\mathrm{d}}}$ and ${\binom{{N-K}}{k_{\mathrm{d}}%
}}\leq (N-K)^{k_{\mathrm{d}}}$ and the summation formula for geometric series, we
obtain
\begin{eqnarray*}
P_{\mathrm{err}} &\leq &\frac{1}{2L}\sum_{i=0}^{L-1}\sum_{k_{\mathrm{d}}=1}^{K}{%
\binom{K}{k_{\mathrm{d}}}}{\binom{{N-K}}{k_{\mathrm{d}}}}\left( \frac{\bar{\lambda}}{4}\right)
^{-\kappa k_{\mathrm{d}}T} \\
&\leq &\frac{1}{2}\sum_{k_{\mathrm{d}}=1}^{K}\left[ \frac{K\left( N-K\right) }{\left(
\bar{\lambda}/4\right) ^{\kappa T}}\right] ^{k_{\mathrm{d}}} \\
&\leq &\frac{1}{2}\frac{\frac{K\left( N-K\right) }{\left( \bar{\lambda}%
/4\right) ^{\kappa T}}}{1-\frac{K\left( N-K\right) }{\left( \bar{\lambda}%
/4\right) ^{\kappa T}}}.\ \ \ \ \blacksquare
\end{eqnarray*}

We make several comments here. First, $\bar{\lambda}$ depends solely on the
measurement matrix $A$. Compared with the results in \cite%
{Wainwright2007Bound}, where the bounds involve the signal, we get more
insight into what quantity of the measurement matrix is important in support
recovery. This information is obtained by modelling the signals $\bs x(t)$
as Gaussian random vectors. The quantity $\bar{\lambda}$ effectively
characterizes system \eqref{model_vector}'s ability to distinguish different
supports. Clearly, $\bar{\lambda}$ is related to the restricted isometry property (RIP), which guarantees stable sparse signal recovery in Compressive Sensing \cite{Candes2005Decoding,Candes2008IntroCS,Baraniuk2007CompressiveSensing}. We discuss the relationship between RIP and $\bar{\lambda}$ for the special case with $K = 1$ at the end of Section \ref{sec:noiseeffectupper}. However, a precise relationship for the general case is yet to be discovered.

Second, we observe that increasing the number of temporal samples plays two
roles simultaneously in the measurement system. For one thing, it decreases
the the threshold $4[K(N-K)]^{\frac{1}{\kappa T}}$ that $\bar{\lambda}$ must
exceed for the bound \eqref{bd_multiple} to hold. However, since $%
\lim_{T\rightarrow \infty }4[K(N-K)]^{\frac{1}{\kappa T}}=4$ for fixed $K$
and $N$, increasing temporal samples can reduce the threshold only to a certain
limit. For another, since the bound \eqref{bd_multiple} is proportional to $e^{-T \log(\bar{\lambda}/4)}$, the probability of error turns to 0 exponentially fast as $T$ increases, as long as $\bar{\lambda}>4\left[ K\left( N-K\right) \right] ^{\frac{1}{\kappa T}}$ is satisfied.

In addition, the final bound \eqref{bd_multiple} is of the same order as the
probability of error when $k_{\mathrm{d}}=1$. The probability of error $P_{\mathrm{err%
}}$ is dominated by the probability of error in cases for which the
estimated support differs by only one index from the true support, which are
the most difficult cases for the decision rule to make a choice. However, in
practice we can imagine that these cases induce the least loss. Therefore,
if we assign weights/costs to the errors based on $k_{\mathrm{d}}$, then the weighted
probability of error or average cost would be much lower. For example, we
can choose the costs to exponentially decrease when $k_{\mathrm{d}}$ increases.
Another possible choice of cost function is to assume zero cost when $k_{\mathrm{d}}$
is below a certain critical number. Our results can be easily extended to
these scenarios.

Finally, note that our bound \eqref{bd_multiple} applies to any
non-degenerate matrix. In Section \ref{sec:gaussian}, we apply the bound to
Gaussian measurement matrices. The additional structure allows us to derive
more profound results on the behavior of the bound.

\subsection{The Effect of
Noise}\label{sec:noiseeffectupper}

In this subsection, we explore how the noise variance affects the
probability of error, which is equivalent to analyzing the behavior of $\bar{%
\lambda}_{S_{i},S_{j}}$ and $\bar{\lambda}$ as indicated in \eqref{bd_binary}
and \eqref{bd_multiple}.

We now derive bounds on the eigenvalues of $H$. The
lower bound is expressed in terms of the QR decomposition of a submatrix of
the measurement matrix with the noise variance $\sigma ^{2}$ isolated.

\begin{pro}
\label{eig_bound} For any non-degenerate measurement matrix $A$, let $%
H=\Sigma _{0}^{1/2}\Sigma _{1}^{-1}\Sigma _{0}^{1/2}$ with $\Sigma
_{i}=A_{S_{i}}A_{S_{i}}^{\dagger }+\sigma ^{2}\eye_{M}$, $k_{\mathrm{i}}=|S_{0}\cap
S_{1}|,k_{0}=\left\vert S_{0}\backslash S_{1}\right\vert =\left\vert
S_{0}\right\vert -k_{\mathrm{i}},k_{1}=\left\vert S_{1}\backslash S_{0}\right\vert
=\left\vert S_{1}\right\vert -k_{\mathrm{i}}$. We have the following:

\begin{enumerate}
\item if $M\geqslant k_{0}+k_{1}$, then the sorted eigenvalues of $H$ that are greater than 1 are lower bounded by the corresponding eigenvalues of \ $\eye_{k_{0}}+%
\frac{1}{\sigma ^{2}}R_{33}R_{33}^{\dagger }$, where $R_{33}$ is the $%
k_{0}\times k_{0}$ submatrix at the lower-right corner of the upper triangle
matrix in the QR decomposition of $%
\begin{bmatrix}
A_{S_{1}\backslash S_{0}} & A_{S_{1}S_{0}} & A_{S_{0}\backslash S_{1}}%
\end{bmatrix};%
$

\item the eigenvalues of $H$ are upper bounded by the corresponding eigenvalues of $\eye_{M}+\frac{1%
}{\sigma ^{2}}A_{S_{0}\backslash S_{1}}A_{S_{0}\backslash S_{1}}^{\dagger }$; in particular, the sorted eigenvalues of $H$ that are greater than $1$ are upper bounded by the corresponding ones of $\eye_{k_0}+\frac{1%
}{\sigma ^{2}}A^{\dagger }_{S_{0}\backslash S_{1}}A_{S_{0}\backslash S_{1}}$.
\end{enumerate}
\end{pro}

\textbf{Proof:} See Appendix B.

\bigskip The importance of this proposition is twofold. First, by isolating the
noise variance from the expression of matrix $H$, this theorem clearly shows
that when noise variance decreases to zero, the relatively large eigenvalues
of $H$ will blow up, which results in increased performance in support
recovery. Second, the bounds provide ways to analyze special measurement
matrices, especially the Gaussian measurement ensemble discussed in Section \ref{sec:gaussian}.

We have the following corollary:
\vspace{-0.2cm}
\begin{cor}
\label{noise_effect} For support recovery problems \eqref{BHT_support} and %
\eqref{MHT_support} with support size $K$, suppose $M\geq 2K$; then there
exist constants $c_{1},c_{2}>0$ that depend only on the measurement matrix $%
A $ such that
\begin{equation}\label{l_bd}
1+\frac{c_{2}}{\sigma ^{2}}\geq \bar{\lambda}\geq 1+\frac{c_{1}}{\sigma ^{2}}%
.
\end{equation}%
From \eqref{bd_binary} and \eqref{bd_multiple}, we then conclude that for any temporal sample size $T$
\begin{equation}
\lim_{\sigma ^{2}\rightarrow 0}P_{\mathrm{err}}=0
\end{equation}%
and the speed of convergence is approximately $(\sigma ^{2})^{\kappa k_{\mathrm{d}}T}$
and $(\sigma ^{2})^{\kappa T}$ for the binary and multiple cases,
respectively.
\end{cor}

\textbf{Proof: }According to Proposition \ref{eig_bound}, for any fixed $%
S_{i},S_{j}$, the eigenvalues of $H=\Sigma _{i}^{1/2}\Sigma
_{j}^{-1}\Sigma
_{i}^{1/2}$ that are greater than $1$ are lower bounded by those of $\eye%
_{k_{\mathrm{d}}}+\frac{1}{\sigma ^{2}}R_{33}R_{33}^{\dagger }$; hence we have
\begin{eqnarray}
\bar{\lambda}_{S_{i},S_{j}} &\geq &\left\vert \eye_{k_{\mathrm{d}}}+\frac{1}{\sigma
^{2}}R_{33}R_{33}^{\dagger }\right\vert ^{1/k_{\mathrm{d}}}  \notag \\
&\geq &\left\vert \eye_{k_{\mathrm{d}}}+\frac{1}{\sigma ^{2}}R_{33}^2\right\vert ^{1/k_{\mathrm{d}}}  \notag \\
&=&\left[ \prod_{l=1}^{k_{\mathrm{d}}}\left( 1+\frac{1}{\sigma ^{2}}r_{ll}^{2}\right) %
\right] ^{1/k_{\mathrm{d}}}  \notag \\
&\geq &1+\frac{1}{\sigma ^{2}}\left( \prod_{l=1}^{k_{\mathrm{d}}}r_{ll}^{2}\right)
^{1/k_{\mathrm{d}}},  \label{bd_l_barlambda}
\end{eqnarray}%
where $r_{ll}$ is the $l$th diagonal element of $R_{33}$. For the second
inequality we have used Fact 8. 11. 20 in \cite{Bernstein2005Matrix}. Since $%
A$ is non-degenerate and $M \geq 2K$, $
\begin{bmatrix}
A_{S_{j}\backslash S_{i}} & A_{S_{j}S_{i}} & A_{S_{i}\backslash S_{j}}%
\end{bmatrix}%
$ is of full rank and $r_{ll}^{2}>0, 0\leq l\leq k_{\mathrm{d}}$ \thinspace for all $%
S_{i}, S_{j}$. Defining $c_{1}$ as the minimal value of $\left(
\prod_{l=1}^{k_{\mathrm{d}}}r_{ll}^{2}\right) ^{1/k_{\mathrm{d}}}$'s over all
possible support pairs $S_{i}, S_{j}$, we then have $c_{1}>0$ and
\begin{equation*}
\bar{\lambda}\geq 1+\frac{c_{1}}{\sigma ^{2}}.
\end{equation*}%
On the other hand, the upper bound on the eigenvalues of $H$ yields
\begin{eqnarray}
\bar{\lambda}_{S_{i},S_{j}} &\leq &\left\vert \eye_{k_{\mathrm{d}}}+\frac{1}{\sigma
^{2}}A^{\dagger
}_{S_{i}\backslash S_{j}}A_{S_{i}\backslash S_{j}}\right\vert ^{1/k_{\mathrm{d}}}  \notag \\
&\leq &1+\frac{1}{\sigma ^{2}k_{\mathrm{d}}}\tr\left( A^{\dagger }_{S_{i}\backslash
S_{j}}A_{S_{i}\backslash Sj}\right)  \notag \\
& = &1+\frac{1}{\sigma ^{2}k_{\mathrm{d}}}\sum_{\substack{ 1\leq m\leq M  \\ n\in
S_{i}\backslash S_{j}}}\left\vert A_{mn}\right\vert ^{2}.
\label{bd_u_barlambda}
\end{eqnarray}%
Therefore, we have
\begin{equation*}
\bar{\lambda}\leq 1+\frac{c_{2}}{\sigma ^{2}},
\end{equation*}%
with $c_{2}=\max_{S:|S|\leq K}\frac{1}{|S|}\sum_{\substack{ 1\leq m\leq M  \\ n\in S}}%
|A_{mn}|^{2}$. All other statements in the theorem follows immediately from %
\eqref{bd_binary} and \eqref{bd_multiple}. $\blacksquare $

\bigskip Corollary \ref{noise_effect} suggests that in the limiting case
where there is no noise, $M\geq 2K$ is sufficient to recover a $K-$sparse
signal. This fact has been observed in \cite{Candes2005Decoding}. Our
result also shows that the optimal decision rule, which is unfortunately
inefficient, is robust to noise. Another extreme case is when the
noise variance $\sigma ^{2}$ is very large. Then from $\log (1+x)\thickapprox
x,0<x<<1$, the bounds in \eqref{bd_binary} and \eqref{bd_multiple} are
approximated by $e^{-\kappa k_{\mathrm{d}}T/\sigma ^{2}}$ and $e^{-\kappa T/\sigma
^{2}}$. Therefore, the convergence exponents for the bounds are proportional to the SNR in this limiting case.

The diagonal elements of $R_{33}$, $r_{ll}$'s, have clear meanings. Since QR
factorization is equivalent to the Gram-Schmidt orthogonalization procedure,
$r_{11}$ is the distance of the first column of $A_{S_{i}/S_{j}}$ to the
subspace spanned by the columns of $A_{S_{j}}$; $r_{22}$ is the distance of
the second column of $A_{S_{i}/S_{j}}$ to the subspace spanned by the
columns of $A_{S_{j}}$ plus the first column of $A_{S_{i}/S_{j}}$, and so
on. Therefore, $\bar{\lambda}_{S_{i},S_{j}}$ is a measure of how well the
columns of $A_{S_{i}/S_{j}}$ can be expressed by the columns of $A_{S_{j}}$,
or, put another way, a measure of the incoherence between the columns of $%
A_{S_{i}}$ and $A_{S_{j}}$. Similarly, $\bar{\lambda}$ is an indicator of
the incoherence of the entire matrix $A$ of order $K$.

To relate $\bar{\lambda}$ with the incoherence, we consider the case with $%
K=1 $ and $\mathbb{F}=\mathbb{R}$. By restricting our attention to matrices
with \emph{unit} columns, the above discussion implies that a better bound is
achieved if the minimal distance of all pairs of column vectors of matrix $A$
is maximized. Finding such a matrix $A$ is equivalent to finding a matrix
with the inner product between columns as large as possible, since the
distance between two unit vectors $u$ and $v$ is $2-2|<u,v>|$ where $%
<u,v>=u^{\prime }v$ is the inner product between $u$ and $v$. For each integer $s$, the RIP  constant $\delta _{s}$ is defined as the smallest number such that \cite{Candes2008IntroCS, Candes2005Decoding}:
\begin{equation}\label{rip}
1-\delta _{s}\leq \frac{\Vert Ax\Vert _{2}^{2}}{\Vert x\Vert _{2}^{2}}\leq
1-\delta _{s},\ \ \ |\mathrm{supp}(x)|= s.
\end{equation}%
A direct computation shows that $\delta _{2}$ is equal to
the minimum of the absolute values of the inner products between all pairs of
columns of $A$. Hence, the requirements of finding the smallest $\delta _{2}$ that satisfies \eqref{rip} and maximizing $\bar{\lambda}$ coincide when $K = 1$. For general $K$, Milenkovic \emph{et.al.} established a relationship between $\delta_2$ and $\delta_{K}$ via Ger\v{s}gorin's disc theorem \cite{Horn1990Matrix} and discussed them as well as some coding theoretic issues in Compressive Sensing context \cite{Milenkovic2009sublinear}.

\section{An Information Theoretic Lower Bound on Probability of
Error}\label{sec:lower}

In this section, we derive an information theoretic lower bound on the
probability of error for \emph{any} decision rule in the multiple support
recovery problem. The main tool is a variant of the well-known Fano's
inequality\cite{Cover2006Information}. In the variant, the average
probability of error in a multiple-hypothesis testing problem is bounded in
terms of the Kullback-Leibler divergence\cite{Kullback1997Information}.
Suppose that we have a random vector or matrix $\bs Y$ with $L$ possible densities $%
f_{0},\ldots ,f_{L-1}$. Denote the average of the Kullback-Leibler
divergence between any pair of densities by
\begin{equation}
\beta =\frac{1}{L^{2}}\sum_{i,j}D_{KL}(f_{i}||f_{j}).  \label{average_KL}
\end{equation}%
Then by Fano's inequality \cite{Yatracos1988Lower},\cite{Wainwright2007Bound}%
, the probability of error \eqref{Perr} for \emph{any} decision rule to
identify the true density is lower bounded by
\begin{equation}
P_{\mathrm{err}}\geq 1-\frac{\beta +\log 2}{\log L}.  \label{Fano}
\end{equation}

Since in the multiple support recovery problem \eqref{MHT_support}, all the
distributions involved are matrix variate Gaussian distributions with zero mean and different variances, we now compute the Kullback-Leibler divergence
between two matrix variate Gaussian distributions. Suppose $f_{i}=\mathbb{F}%
\mathcal{N}_{M,T}(0,\Sigma _{i}\otimes I_{T}),f_{j}=\mathbb{F}\mathcal{N}%
_{M,T}(0,\Sigma _{j}\otimes I_{T})$, the Kullback-Leibler divergence
has closed form expression:
\begin{eqnarray*}
&&D_{KL}(f_{i}||f_{j})=\mathbb{E}_{f_{i}}\log \frac{f_{i}}{f_{j}} \\
&=&\frac{1}{2}\ \mathbb{E}_{f_{i}}\left[ -\kappa \func{tr}\left[ \boldsymbol{Y}%
^{\dagger }\left( \Sigma _{i}^{-1}-\Sigma _{j}^{-1}\right) \boldsymbol{Y}%
\right] -\kappa T\log \frac{\left\vert \Sigma _{i}\right\vert }{\left\vert
\Sigma _{j}\right\vert }\right] \\
&=&\frac{1}{2}\kappa T\left[\func{tr}\left( H_{i,j}-\eye_{M}\right)
+\log \frac{\left\vert \Sigma _{j}\right\vert }{\left\vert \Sigma
_{i}\right\vert }\right],
\end{eqnarray*}%
where $H_{i,j}=\Sigma _{i}^{1/2}\Sigma _{j}^{-1}\Sigma _{i}^{1/2}$.
Therefore, we obtain the average Kullback-Leibler divergence %
\eqref{average_KL} for the multiple support recovery problem as
\begin{eqnarray*}
\beta &=&\frac{1}{L^{2}}\sum_{S_i, S_j}\frac{1}{2}\kappa T\left[\mathrm{\func{tr}}%
(H_{i,j})-M + \log \frac{|%
{\Sigma }_{j}|}{|{\Sigma }_{i}|}\right] \\
&=&\frac{\kappa T}{2L^{2}}\sum_{S_i,S_j}\left[\mathrm{\func{tr}}(H_{i,j})-M%
\right],
\end{eqnarray*}%
where the $\log \frac{|{\Sigma
}_{j}|}{|{\Sigma }_{i}|}$ terms all cancel out and $L = {N \choose K}$.
Invoking the second part of Proposition \ref{eig_bound}, we get
\begin{eqnarray*}
\mathrm{\func{tr}}\left( H_{i,j}\right)
&\leq & \mathrm{tr}\left(\eye_M + \frac{1}{\sigma^2}A_{S_i\backslash S_j}A^\dagger_{S_i\backslash S_j}\right)\\
& = & M + \frac{1}{\sigma^2} \sum_{_{\substack{ 1\leq m\leq M  \\ n\in
S_{i}\backslash S_{j}}}}\left\vert A_{mn}\right\vert ^{2}.
\end{eqnarray*}%
Therefore, the average Kullback-Leibler divergence is
bounded by
\begin{equation*}
\beta \leq \frac{\kappa T}{2\sigma^2L^{2}}\sum_{S_i, S_j}\sum_{_{\substack{ 1\leq m\leq M  \\ n\in
S_{i}\backslash S_{j}}}}\left\vert A_{mn}\right\vert ^{2}.
\end{equation*}

Due to the symmetry of the right-hand side, it must be of the form $a
\sum_{_{\substack{ 1\leq m\leq M  \\ 1\leq n\leq N}}}\left\vert
A_{mn}\right\vert ^{2}=a \left\Vert A\right\Vert _{\mathrm{F}}^{2}$, where $%
\Vert \cdot \Vert _{\mathrm{F}}$ is the Frobenius norm. Setting all $A_{mn}=1$ gives
\begin{eqnarray*}
&&  \frac{\kappa T}{2\sigma^2L^{2}}\sum_{S_i, S_j}\sum_{_{\substack{ 1\leq m\leq M  \\ n\in
S_{i}\backslash S_{j}}}}1 \\
&=& \frac{\kappa T}{2\sigma^2L^{2}}\sum_{i = 0}^{L-1} \sum_{k_{\mathrm d} = 1}^K \binom{K}{k_{\mathrm{d}}}\binom{N-K}{k_{\mathrm{d}}}
k_{\mathrm{d}}M \\
& = & a MN.
\end{eqnarray*}
Therefore, we get $a =\frac{\kappa TK\left( N-K\right) }{2\sigma ^{2}N^{2}}$ using the mean expression for hypergeometric distribution:
\begin{equation*}
\sum_{k_{\mathrm{d}}=1}^{K}\frac{\binom{K}{k_{\mathrm{d}}}\binom{N-K}{k_{\mathrm{d}}}}{\binom{N}{K}}%
k_{\mathrm{d}}=\frac{K\left( N-K\right) }{N}.
\end{equation*}%
Hence, we have
\begin{equation*}
\beta \leq \frac{\kappa TK\left( N-K\right) }{2\sigma ^{2}N^{2}}\left\Vert
A\right\Vert _{\mathrm{F}}^{2}.
\end{equation*}%
Therefore, the probability of error is lower bounded by
\begin{equation}
P_{\mathrm{err}}\geq 1-\frac{\frac{\kappa TK\left( N-K\right) }{2\sigma
^{2}N^{2}}\left\Vert A\right\Vert _{\mathrm{F}}^{2}+\log 2}{\log L}.
\end{equation}%
We conclude with the following theorem:

\begin{thm}\label{thm_bd_lower}
For multiple support recovery problem \eqref{MHT_support}, the probability
of error for any decision rule is lower bounded by%
\begin{equation}
P_{\mathrm{err}}\geq 1-\frac{\kappa T\frac{K}{N}\left( 1-\frac{K}{N}\right)
\left\Vert A\right\Vert _{\mathrm{F}}^{2}}{2\sigma ^{2}\log \binom{N}{K}}+o\left(
1\right) .  \label{bd_lower}
\end{equation}
\end{thm}

Each term in bound $\eqref{bd_lower}$ has clear meanings. The Frobenius norm
of measurement matrix $\left\Vert A\right\Vert _{\mathrm{F}}^{2}$ is total gain of system \eqref{model_scalar}. Since the measured
signal is $K-$sparse, only a fraction of the gain plays a role in the
measurement, and its average over all possible $K-$sparse signals is $\frac{K%
}{N}\left\Vert A\right\Vert _{\mathrm{F}}^{2}$. While an increase in signal energy enlarges the distances between signals, a penalty term $\left( 1-\frac{K}{N}\right)$ is introduced because we now have more signals. The term $%
\log L=\log \binom{N}{K}$ is the total uncertainty or entropy of the support
variable $\bs S,$ since we impose a uniform prior on it. As long as $K\leq \frac{%
N}{2}$, increasing $K$ increases both the average gain exploited by the
measurement system, and the entropy of the support variable $%
\bs S $. The overall effect, quite counterintuitively, is a decrease of the
lower bound in \eqref{bd_lower}. Actually, the term involving $K$, $\frac{\frac{K}{N}(1-\frac{K}{N})}{\log {N \choose K}}$, is approximated by an increasing function $\frac{\alpha(1-\alpha)}{N H(\alpha)}$ with $\alpha = \frac{K}{N}$ and the binary entropy function $H(\alpha) = -\alpha\log \alpha - (1-\alpha)\log(1-\alpha)$. The reason for the decrease of the bound is that the bound only involves the \emph{effective} SNR without regard to any inner structure of $A$ (e.g. the incoherence) and the effective SNR increases with $K$. To see this, we compute the effective SNR as $\mathbb{E}_{\bs x} \left[\frac{\|A\bs x\|_{\mathrm{F}}^2}{M\sigma^2}\right]=\frac{\frac{K}{N}\|A\|_{\mathrm{F}}^2}{M\sigma^2}$. If we scale down the effective SNR through increasing the noise energy $\sigma^2$ by a factor of $K$, then the bound is strictly increasing with $K$.

The above analysis suggests that the lower bound \eqref{bd_lower} is weak as it disregards any incoherence property of the measurement matrix $A$. For some cases, the bound reduces to $2\kappa \sigma^2 \log \frac{N}{K}$ (refer to Corollary \ref{snet}, Theorem \ref{DOA} and \ref{Gaussian_nece}) and is less than $K$ when the noise level or $K$ is relatively large. Certainly recovering the support is not possible with fewer than $K$ measurements. The bound is loose also in the sense that when $T$, $%
\left\Vert A\right\Vert _{\mathrm{F}}^{2}$, or the SNR $1/
\sigma^2$ is large enough the bound
becomes negative, but when there is noise, perfect support recovery is\
generally impossible. While the original Fano's inequality
\begin{equation}
H\left( P_{\mathrm{err}}\right) +P_{\mathrm{err}}\log \left( L-1\right) \geq
H\left( \boldsymbol{S}|\boldsymbol{Y}\right) \label{fano_original}
\end{equation}%
is tight in some cases\cite{Cover2006Information}, the adoption of the average divergence %
\eqref{average_KL} as an upper bound on the mutual information $I\left(
\boldsymbol{S};\boldsymbol{Y}\right) $ between the random support $%
\boldsymbol{S}$ and the observation $\boldsymbol{Y}$ reduces the tightness (see the proof of \eqref{Fano} in \cite%
{Birg1983Approximation}). Due to the difficulty of computing $H(\bs S|\bs Y)$ and $I(\bs S; \bs Y)$ analytically, it is not clear whether a direction application of \eqref{fano_original} results in a significantly better bound.

Despite of its drawbacks we discussed, the bound \eqref{bd_lower} identifies the importance of the gain $\|A\|_{\mathrm{F}}^2$ of the measurement matrix, a quantity usually ignored in, for example, Compressive Sensing. We can also draw some interesting conclusions from \eqref{bd_lower} for measurement matrices with special properties. In particular, in the following corollary, we consider measurement matrices with rows or columns normalized to one. The rows of a measurement matrix are normalized to one in sensor network scenario (SNET) where each sensor is power limited while the columns are sometimes normalized to one in Compressive Sensing (refer to \cite{Aeron2009fundamental} and references therein).

\begin{cor}\label{snet}
In order to have a probability of error $P_{\mathrm{err}} < \varepsilon$ with $0 < \varepsilon < 1$, the number of measurements must satisfy:
\begin{eqnarray}
  MT &\geq & (1-\varepsilon)\frac{2\sigma^2 K \log \frac{N}{K}}{\kappa \frac{K}{N}(1-\frac{K}{N})} + o(1) \nonumber \\
  &\geq &  (1-\varepsilon)\frac{8\sigma^2}{\kappa} K \log \frac{N}{K} + o(1), \label{bd_row_unit}
\end{eqnarray}
if the rows of $A$ have unit norm; and
\begin{eqnarray}
  T &\geq & (1-\varepsilon)\frac{2\sigma^2 \log \frac{N}{K}}{\kappa \left(1 - \frac{K}{N}\right)} + o(1) \nonumber \\
  &\geq & (1 - \varepsilon)\frac{2\sigma^2}{\kappa} \log \frac{N}{K} + o(1), \label{bd_col_unit}
\end{eqnarray}
if the columns of $A$ have unit norm.
\end{cor}

Note that the necessary condition \eqref{bd_row_unit} has the same critical quantity as the sufficient condition in Compressive Sensing. The inequality in \eqref{bd_col_unit} is independent of $M$. Therefore, if the columns are normalized to have unit norm, it is necessary to have multiple temporal measurements for a vanishing probability of error. Refer to Theorem \ref{DOA} and \ref{Gaussian_nece} and discussions following them.

In the work of \cite{Willsky2005Source}, each column of $A$ is the
array manifold vector function evaluated at a sample of the
DOA parameter. The implication of the bound \eqref{bd_lower} for
optimal design is that we should construct an array whose geometry
leads to maximal $\left\Vert A\right\Vert _{\mathrm{F}}^{2}$. However,
under the narrowband signal assumption and narrowband array
assumption \cite{Dogandzic2002Space}, the array manifold
vector for isotropic sensor arrays always has norm $\sqrt{M}$\cite%
{Vantrees2002Optimum}, which means that $\left\Vert A\right\Vert _{\mathrm{F}}^{2}=MN$%
. Hence in this case, the probability of error is always bounded by
\begin{equation}
P_{\mathrm{err}}\geq 1-\frac{T\frac{K}{N}\left( 1-\frac{K}{N}\right)
MN}{2\sigma ^{2}\log \binom{N}{K}}+o\left( 1\right) .
\end{equation}%
Therefore, we have the following theorem,

\begin{thm}
\label{DOA} Under the narrowband signal assumption and narrowband array
assumption, for an isotropic sensor array in the DOA estimation scheme
proposed in \cite{Willsky2005Source}, in order to
let the probability of error $P_{\mathrm{err}}<\varepsilon $ with $0 <
\varepsilon < 1$ for any decision rule, the number of measurements must
satisfy the following:
\begin{eqnarray}
MT &\geq &\left( 1-\varepsilon \right) \frac{2\sigma ^{2}\log \binom{N}{K}}{%
K\left( 1-\frac{K}{N}\right) }+o(1) \nonumber \\
&\geq &\left( 1-\varepsilon \right) 2\sigma ^{2}\log \frac{N}{K}+o(1).
\end{eqnarray}
\end{thm}

We comment that the same lower bound applies to Fourier
measurement matrix (not normalized by $1/\sqrt{M}$) due to the
same line of argument. We will not explicitly present this result
in the current paper.

Since in radar and sonar applications the number of targets $K$ is usually small, our result shows that the number of samples is
lower bounded by $\log N$. Note that $N$ is the
number of intervals we use to divide the whole range of DOA;
hence, it is a measure of resolution. Therefore, the number of
samples only needs to increase in the logarithm of $N$, which is
very desirable. The symmetric roles played by $M$ and $T$ are also
desirable since $M$ is the number of sensors and is expensive to
increase. As a consequence, we simply increase the number of
samples to achieve a desired probability of error. In addition, unlike the upper bound of Theorem \ref{thm_mul_bd}, we do not need to assume that $M \geq 2K$ in Theorem \ref{thm_bd_lower} and \ref{DOA}. Actually, Malioutov \emph{et.al.} made the empirical observation that $\ell_1-$SVD technique can resolve $M-1$ sources if they are well separated \cite{Willsky2005Source}. Theorem \ref{DOA} still applies to this extreme case.

Analysis of support recovery problem with measurement matrix obtained from sampling a manifold has considerable complexity compared with the Gaussian case. For example, it presents significant challenge to estimate $\bar{\lambda}_{S_i,S_j}$ in the DOA problem except for a few special cases that we discuss in \cite{Tang2010DOA}. As we mentioned before, unlike the Gaussian case, $\bar{\lambda}_{S_i,S_j}$ for uniform linear arrays varies greatly with $S_i$ and $S_j$. Therefore, even if we can compute $\bar{\lambda}_{S_i,S_j}$, replacing it with $\bar{\lambda}$ in the upper bound of Theorem \ref{thm_mul_bd} would lead to a very loose bound. On the other hand, the lower bound of Theorem \ref{thm_bd_lower} only involves the Frobenius norm of the measurement matrix, so we apply it to the DOA problem effortlessly. However, the lower bound is weak as it does not exploit any inner structure of the measurement matrix.

Donoho \emph{et.al.} considered the recovery of a ``sparse" wide-band signal from narrow-band measurements \cite{donoho1989uncertainty, donoho1992sieve}, a problem with essentially the same mathematical structure when we sample the array manifold uniformly in the wave number domain instead of the DOA domain. It was found that the spectral norm of the product of the band-limiting and time-limiting operators is crucial to stable signal recovery measured by the $l_2$ norm. In \cite{donoho1989uncertainty}, Donoho and Stark bounded the spectral norm using the Frobenius norm, which leads to the well-known uncertainty principle. The authors commented that the uncertainty principle condition demands an extreme degree of sparsity for the signal. However, this condition can be relaxed if the signal support are widely scattered. In \cite{Willsky2005Source}, Malioutov \emph{et.al.} also observed from numerical simulations that the $\ell_1-$SVD algorithm performs much better when the sources are well separated than when they are located close together. In particular, they observed that presence of bias is mitigated greatly when sources are far apart. Donoho and Logan \cite{donoho1992sieve} explored the effect of the scattering of the signal support by using the ``analytic principle of the large sieve". They bounded the spectral norm for the limiting operator by the maximum Nyquist density, a quantity that measures the degree of scattering of the signal support. We expect that our results can be improved in a similar manner. The challenges include using support recovery as a performance measure, incorporating multiple measurements, as well as developing the whole theory within a probabilistic framework.

\section{Support Recovery for the Gaussian Measurement Ensemble}

\label{sec:gaussian}

In this section, we refine our results in previous sections from general
non-degenerate measurement matrices to the Gaussian ensemble. Unless
otherwise specified, we always assume that the elements in a measurement
matrix $\bs A$ are \emph{i.i.d.} samples from unit variance real or complex
Gaussian distributions. The Gaussian measurement ensemble is widely used and
studied in Compressive Sensing \cite%
{Candes2006Uncertainty,Donoho2006Compressed,Candes2005Decoding,Candes2008IntroCS,Baraniuk2007CompressiveSensing}%
. The additional structure and the theoretical tools available enable us to
derive deeper results in this case. In this section, we assume general scaling of $(N, M, K, T)$. We do not find  in our results a clear distinction between the regime of sublinear sparsity and the regime of linear sparsity as the one discussed in \cite{Wainwright2007Bound}.

We first show two corollaries on the
eigenvalue structure for the Gaussian measurement ensemble. Then we derive
sufficient and necessary conditions in terms of $M,N,K$ and $T$ for the
system to have a vanishing probability of error.

\subsection{Eigenvalue Structure for a Gaussian Measurement Matrix}
\label{sec:gaussian_eigen}
First, we observe that a Gaussian measurement matrix is non-degenerate with
probability one, since any $p\leq M$ random vectors $\bs a_{1},\bs %
a_{2},\ldots ,\bs a_{p}$ from $\mathbb{F}\mathcal{N}(0,\Sigma )$ with $%
\Sigma \in \mathbb{R}^{M\times M}$ positive definite are linearly
independent with probability one (refer to Theorem 3.2.1 in \cite%
{Gupta1999Matrix}). As a consequence, we have

\begin{cor}
\label{eig_count_Gaussian} For Gaussian measurement matrix $\bs A$, let $%
\boldsymbol{H}=\boldsymbol{\Sigma }_{0}^{1/2}\boldsymbol{\Sigma }_{1}^{-1}%
\boldsymbol{\Sigma }_{0}^{1/2}$, $k_{\mathrm{i}}=|S_{0}\cap S_{1}|,k_{0}=\left\vert
S_{0}\backslash S_{1}\right\vert =\left\vert S_{0}\right\vert
-k_{\mathrm{i}},k_{1}=\left\vert S_{1}\backslash S_{0}\right\vert =\left\vert
S_{1}\right\vert -k_{\mathrm{i}}$. If $M\geqslant k_{0}+k_{1}$, then with
probability one,  $k_{0}$ eigenvalues of matrix $\boldsymbol{H}$ are greater
than $1$, $k_{1}$ less than $1$, and $M-\left( k_{0}+k_{1}\right) $ equal to
$1$.
\end{cor}

We refine Proposition \ref{eig_bound} based on the well-known QR
factorization for Gaussian matrices
\cite{Gupta1999Matrix},\cite{Anderson2003Multivariate}.

\begin{cor}
With the same notations as in Corollary \ref{eig_count_Gaussian}, then
with probability one, we have:
\begin{enumerate}
\item if $M\geqslant k_{0}+k_{1}$, then the sorted eigenvalues of $\bs H$ that are greater than 1 are lower bounded by the corresponding ones of \ $\eye_{k_{0}}+\frac{1}{%
\sigma ^{2}}\boldsymbol{R}_{33}\boldsymbol{R}_{33}^{\dagger }$, where the
elements of $\boldsymbol{R}_{33}=\left( r_{mn}\right) _{k_{0}\times k_{0}}$
satisfy:
\begin{eqnarray*}
2\kappa r_{mn}^{2} &\sim &\chi _{2\kappa (M-k_{1}-k_{\mathrm{i}}-m+1)}^{2},\ \ 1 \leq m \leq k_0, \\
r_{mn} &\sim &\mathbb{F}\mathcal{N}\left( 0,1\right) ,\ \ 1\leq m<n\leq k_{0}.
\end{eqnarray*}

\item the eigenvalues of $\bs H$ are upper bounded by the corresponding eigenvalues of $\eye_{M}+\frac{1%
}{\sigma ^{2}}\bs{A}_{S_{0}\backslash S_{1}}\bs{A}_{S_{0}\backslash S_{1}}^{\dagger }$; in particular, the sorted eigenvalues of $\bs H$ that are greater than $1$ are upper bounded by the corresponding ones of $\eye_{k_0}+\frac{1%
}{\sigma ^{2}}\bs{A}^{\dagger }_{S_{0}\backslash S_{1}}\bs{A}_{S_{0}\backslash S_{1}}$.
\end{enumerate}
\end{cor}

Now with the distributions on the elements of the bounding matrices, we can
give sharp estimate on $\bar{\lambda}_{S_{i},S_{j}}$. In particular, we have the following proposition:

\begin{pro}
For Gaussian measurement matrix $\bs A$, suppose $S_{i}$ and $S_{j}$ are a
pair of distinct supports with the same size $K$. Then we have%
\begin{eqnarray*}
1+\frac{M}{\sigma ^{2}} &\geq &\mathbb{E}\bar{\lambda}_{S_{i},S_{j}}\geq 1+%
\frac{M-K-k_{\mathrm{d}}}{\sigma ^{2}}.
\end{eqnarray*}
\end{pro}

\textbf{Proof: } We copy the inequalities \eqref{bd_l_barlambda}, %
\eqref{bd_u_barlambda} on $\bar{\lambda}_{S_{i},S_{j}}$ here:

\begin{equation*}
1+\frac{1}{\sigma ^{2}k_{\mathrm{d}}}\sum_{\substack{ 1\leq m\leq M  \\ n\in
S_{i}\backslash S_{j}}}\left\vert A_{mn}\right\vert ^{2}\geq \bar{\lambda}%
_{S_{i},S_{j}}\geq 1+\frac{1}{\sigma ^{2}}\left(
\prod_{m=1}^{k_{\mathrm{d}}}|r_{mm}|^{2}\right) ^{1/k_{\mathrm{d}}}.
\end{equation*}

The proof then
reduces to the computation of two expectations, one of which is trivial:%
\begin{eqnarray*}
&&\mathbb{E}\frac{1}{\sigma ^{2}k_{\mathrm{d}}}\sum_{\substack{ 1\leq m\leq M  \\ %
n\in S_{0}\backslash S_{1}}}\left\vert \boldsymbol{A}_{mn}\right\vert ^{2}=%
\frac{M}{\sigma ^{2}}.
\end{eqnarray*}

Next, the independence of the $r_{nn}$'s and the convexity of exponential
functions together with Jensen's inequality yield
\begin{eqnarray*}
&&\mathbb{E}\frac{1}{\sigma ^{2}}\left( \prod_{n=1}^{k_{\mathrm{d}}}r_{nn}^{2}\right)
^{1/k_{\mathrm{d}}} \\
&=&\frac{1}{2\kappa \sigma ^{2}}\mathbb{E}\exp \left[ \frac{1}{k_{\mathrm{d}}}%
\sum_{n=1}^{k_{\mathrm{d}}}\log \left( 2\kappa r_{nn}^{2}\right) \right] \\
&\geq &\frac{1}{2\kappa \sigma ^{2}}\exp \left[ \frac{1}{k_{\mathrm{d}}}%
\sum_{n=1}^{k_{\mathrm{d}}}\mathbb{E}\log \left( 2\kappa r_{nn}^{2}\right) \right] .
\end{eqnarray*}

Since $\left( 2\kappa r_{nn}^{2}\right) \sim \chi _{2\kappa (M-K-n+1)}^{2}$,
the expectation of logarithm is $\mathbb{E}\log \left( 2\kappa r_{nn}^{2}\right) =\log
2+\psi \left( \kappa (M-K-n+1)\right) $, where $\psi \left( z\right) =\frac{%
\Gamma ^{\prime }\left( z\right) }{\Gamma \left( z\right) }$ is the digamma
function. Note that $\psi \left( z\right) $ is increasing and satisfies $%
\psi \left( z+1\right) \geq \log z$. Therefore, we have%
\begin{eqnarray*}
&&\mathbb{E}\frac{1}{\sigma ^{2}}\left( \prod_{n=1}^{k_{\mathrm{d}}}r_{nn}^{2}\right)
^{1/k_{\mathrm{d}}} \\
&\geq &\frac{1}{2\kappa \sigma ^{2}}\exp \left[ \log 2+\frac{1}{k_{\mathrm{d}}}%
\sum_{n=1}^{k_{\mathrm{d}}}\psi \left( \kappa (M-K-n+1)\right) \right] \\
&\geq &\frac{1}{\kappa \sigma ^{2}}\exp \left[ \psi \left( \kappa
(M-K-k_{\mathrm{d}}+1)\right) \right] \\
&\geq &\frac{1}{\kappa \sigma ^{2}}\exp \left[ \log \left( \kappa
(M-K-k_{\mathrm{d}})\right) \right] \\
&\geq &\frac{M-K-k_{\mathrm{d}}}{\sigma ^{2}}.\ \ \ \ \ \ \ \ \ \ \ \ \ \ \ \ \ \ \ \
\ \ \ \ \ \ \ \ \ \blacksquare
\end{eqnarray*}

\bigskip

The expected value of the critical quantity $\bar{\lambda}_{S_{i},S_{j}}$ lies between $1+\frac{M-2K}{\sigma ^{2}}$
and $1+\frac{M}{\sigma ^{2}}$, linearly proportional to $M$. Note that in
conventional Compressive Sensing, the variance of the elements of $%
\boldsymbol{A}$ is usually taken to be $\frac{1}{M}$, which is equivalent to
scaling the noise variance $\sigma ^{2}$ to $M\sigma ^{2}$ in our model. The
resultant $\bar{\lambda}_{S_{i},S_{j}}$ is then centered between $1+\frac{1-2\frac{K}{M}}{\sigma ^{2}}$ and $1+%
\frac{1}{\sigma ^{2}}$.

\subsection{Necessary Condition}\label{sec_gaussian_nece}
\label{sec:gaussian_nece}
One fundamental problem in Compressive Sensing is how many samples should
the system take to guarantee a stable reconstruction. Although many
sufficient conditions are available, non-trivial necessary conditions are
rare. Besides, in previous works, stable reconstruction has been measured in
the sense of $l_{p}$ norms between the reconstructed signal and the true
signal. In this section, we derive two necessary conditions on $M$ and $T$
in terms of $N$ and $K$ in order to guarantee respectively that, first, $%
\mathbb{E}P_{\mathrm{err}}$ turns to zeros and, second, for majority
realizations of $\boldsymbol{A}$, the probability of error vanishes. More
precisely, we have the following theorem:

\begin{thm}\label{Gaussian_nece}
\bigskip In the support recovery problem \eqref{MHT_support}, for any $%
\varepsilon ,\delta >0$, a necessary condition of $\mathbb{E}P_{\mathrm{err}%
}<\varepsilon $ is
\begin{eqnarray}
MT  &\geq &\left( 1-\varepsilon \right) \frac{2\sigma ^{2}\log
\binom{N}{K}}{\kappa K\left( 1-\frac{K}{N}\right) }+o(1)  \label{necessary_mean_1}
\\
&\geq &\left( 1-\varepsilon \right) \frac{2\sigma^2}{\kappa}\log \frac{N}{K}+o(1) ,
\label{necessary_mean_2}
\end{eqnarray}%
and a necessary condition of $\Pr \left\{ P_{\mathrm{err}}\left( \boldsymbol{%
A}\right) \leq \varepsilon \right\} \geq 1-\delta $ is
\begin{eqnarray}
MT &\geq &\left( 1-\varepsilon -\delta \right)
\frac{2\sigma^2 \log \binom{N}{K}}{\kappa K\left( 1-\frac{K}{N}\right) }+o(1)
\label{necessary_prob_1} \\
&\geq &\left( 1-\varepsilon -\delta \right) \frac{2\sigma^2}{\kappa}\log \frac{N}{K}+o(1) .
\label{necessary_prob_2}
\end{eqnarray}
\end{thm}

\textbf{Proof: }Equation \eqref{bd_lower} and $\mathbb{E}\left\Vert
A\right\Vert _{\mathrm{F}}^{2}=\sum_{m,l}\mathbb{E}\left\vert \boldsymbol{A}%
_{ml}\right\vert ^{2}=MN$ give
\begin{equation}\label{average_bound}
\mathbb{E}P_{\mathrm{err}}\geq 1-\frac{\kappa T\frac{K}{N}\left( 1-\frac{K}{N%
}\right) MN}{2\sigma ^{2}\log \binom{N}{K}}+o\left( 1\right) .
\end{equation}%
Hence, $\mathbb{E}P_{\mathrm{err}}<\varepsilon $ entails \eqref{necessary_mean_1} and \eqref{necessary_mean_2}.

Denote by $E$ the event $\left\{ \boldsymbol{A}:P_{\mathrm{err}}\left(
\boldsymbol{A}\right) \leq \varepsilon \right\} $; then $\Pr \left\{
E^{c}\right\} \leq \delta $ and we have%
\begin{eqnarray*}
\mathbb{E}P_{\mathrm{err}} &=&\int_{E}P_{\mathrm{err}}\left( \boldsymbol{A}%
\right) +\int_{E^{c}}P_{\mathrm{err}}\left( A\right)  \\
&\leq &\varepsilon \Pr \left( E\right) +\Pr \left( E^{c}\right)  \\
&\leq &\varepsilon +\delta .
\end{eqnarray*}%
Therefore, from the first part of the theorem, we obtain \eqref{necessary_prob_1} and \eqref{necessary_prob_2}.\ \ \ \ \ \ \ \ \ \ \ $\blacksquare$

\bigskip

We compare our results with those of \cite{Wainwright2007Bound} and \cite{Aeron2009fundamental}. As we mentioned in the introduction, the differences in problem modeling and the definition of the probability of error make a direct comparison difficult. We first note that Theorem 2 in \cite{Wainwright2007Bound} is established for the restricted problem where it is known \emph{a priori} that all non-zero components in the sparse signal are equal. Because the set of signal realizations with equal non-zero components is a rare event in our signal model, it is not fitting to compare our result with the corresponding one in \cite{Wainwright2007Bound} by computing the distribution of the smallest on-support element, e.g., the expectation. Actually, the square of the smallest on-support element for the restricted problem, $\mathcal{M}^2(\beta^*)$ (or $\beta$ in \cite{Aeron2009fundamental}), is equivalent to the signal variance in our model: both are measures of the signal energy. If we take into account the noise variance and replace $\mathcal{M}^2(\beta^*)$ (or $\beta^2\mathrm{SNR}$ in \cite{Aeron2009fundamental}) with $1/\sigma^2$ , the necessary conditions in these papers coincide with ours when only one temporal sample is available.

Our result shows that as far as support recovery is concerned, one
cannot avoid the $\log \frac{N}{K}$ term when only given one
temporal sample. Worse, for conventional Compressive Sensing with
a measurement matrix generated from a Gaussian random variable
with variance $1/M$, the necessary condition becomes

\begin{eqnarray*}
T &\geq &\frac{2\sigma ^{2}\log \binom{N}{K}}{\kappa K\left( 1-\frac{K}{N}%
\right) }+o(1) \\
&\geq &\frac{2\sigma ^{2}}{\kappa }\log \frac{N}{K}+o(1),
\end{eqnarray*}%
which is independent of $M$. Therefore, when there is \emph{considerable} noise \mbox{( $\sigma^2 > \kappa/(2\log \frac{N}{K})$ )}, it is impossible to have a
vanishing $\mathbb{E}P_{\mathrm{err}}$ no matter how large an $M$
one takes. Basically this situation arises because while taking
more samples, one scales down the measurement gains $A_{ml}$,
which effectively reduces the SNR and thus is not helpful in
support recovery. As discussed below Theorem \ref{DOA}, $\log
\binom{N}{K}$ is the uncertainty of the support variable $S$, and $\log \frac{%
N}{K}$ actually comes from it. Therefore, it is no surprise that the number
of samples is determined by this quantity and cannot be made independent of
it.

\subsection{Sufficient Condition}\label{sec_gaussian_suff}
\label{sec:gaussian_suff}
We derive a sufficient condition in parallel with sufficient conditions in
Compressive Sensing. In Compressive Sensing, when only one temporal sample
is available, $M=\Omega \left( K\log \frac{N}{K}\right) $ is enough for
stable signal reconstruction for the majority of the realizations of
measurement matrix $\boldsymbol A$ from a Gaussian ensemble with variance $\frac{1}{M}$.
As shown in the previous subsection, if we take the probability of error for
support recovery as a performance measure, it is impossible in this case to
recover the support with a vanishing probability of error unless the noise is small. Therefore, we
consider a Gaussian ensemble with unit variance. We first establish a lemma
to estimate the lower tail of the distribution for $\bar{\lambda}%
_{S_{i},S_{j}}$. We have shown that the $\mathbb{E}\left( \bar{\lambda}%
_{S_{i},S_{j}}\right) $ lie between $1+\frac{M-2K}{\sigma ^{2}}$ and $1+%
\frac{M}{\sigma ^{2}}$. When $\gamma $ is much less than $1+\frac{M-2K}{%
\sigma ^{2}}$, we expect that $\Pr \left\{ \bar{\lambda}_{S_{i},S_{j}}\leq
\gamma \right\} $ decays quickly. More specifically, we have the following
large deviation lemma:

\begin{lm}
\label{tailprob} Suppose that $\gamma =\frac{1}{3}\frac{M-2K}{\sigma ^{2}}$.
Then there exists constant $c>0$ such that for $M-2K$ sufficiently large, we
have
\begin{equation*}
\Pr \left\{ \bar{\lambda}_{S_{i},S_{j}}\leq \gamma \right\} \leq \exp \left[
-c\left( M-2K\right) \right] .
\end{equation*}
\end{lm}

This large deviation lemma together with the union bound yield the
following sufficient condition for support recovery:

\begin{thm}\label{thm_suff_Gaussian}
Suppose that
\begin{equation}
M=\Omega \left( K\log \frac{N}{K}\right)   \label{sufficient_M}
\end{equation}%
and
\begin{equation}
\kappa T\log \frac{M}{\sigma ^{2}}\gg \log \left[ K\left( N-K\right) \right] .
\label{sufficient_TlogM}
\end{equation}%
Then given any realization of measurement matrix $\boldsymbol{A}$ from a
Gaussian ensemble, the optimal decision rule \eqref{opt_MHT} for multiple
support recovery problem \eqref{MHT_support} has a vanishing $P_{\mathrm{err}%
}$ with probability turning to one. In particular, if $M=\Omega \left( K\log
\frac{N}{K}\right) $ and
\begin{equation}
T\gg \frac{\log N}{\log \log N},  \label{sufficient_T}
\end{equation}%
then the probability of error turns to zero as $N$ turns to infinity.
\end{thm}

\textbf{Proof: }Denote $\gamma =\frac{1}{3}\frac{M-2K}{\sigma ^{2}}$. Then
according to the union bound, we have
\begin{eqnarray*}
&&\Pr \left\{ \bar{\lambda}\leq \gamma \right\}  \\
&=&\Pr \left\{ \bigcup_{S_{i}\neq S_{j}}\left[ \bar{\lambda}%
_{S_{i},S_{j}}\leq \gamma \right] \right\}  \\
&\leq &\sum_{S_{i}\neq S_{j}}\Pr \left\{ \bar{\lambda}_{S_{i},S_{j}}\leq
\gamma \right\} .
\end{eqnarray*}%
Therefore, application of Lemma \ref{tailprob} gives
\begin{eqnarray*}
&&\Pr \left\{ \bar{\lambda}\leq \gamma \right\}  \\
&\leq &\binom{N}{K}^{2}K\exp \left\{ -c\left( M-2K\right) \right\}  \\
&\leq &\exp \left[ -c\left( M-2K\right) +2K\log \frac{N}{K}+\log K\right] .
\end{eqnarray*}%
Hence, as long as $M=\Omega \left( K\log \frac{N}{K}\right) $, we know that
the exponent turns to $-\infty $ as $N\longrightarrow \infty $. We now
define $E=\left\{ \boldsymbol{A}:\bar{\lambda}\left( \boldsymbol{A}\right)
>\gamma \right\} $, where $\Pr \left\{ E\right\} $ approaches one as $N$
turns to infinity. Now the upper bound \eqref{bd_multiple} becomes
\begin{eqnarray*}
P_{\mathrm{err}} &=&O\left( \frac{K\left( N-K\right) }{\left( \frac{\bar{%
\lambda}}{12\sigma ^{2}}\right) ^{\kappa T}}\right)  \\
&=&O\left( \frac{K\left( N-K\right) }{\left( \frac{M}{\sigma ^{2}}\right)
^{\kappa T}}\right) .
\end{eqnarray*}%
Hence, if $\kappa T\log \frac{M}{\sigma ^{2}}\gg \log \left[ K\left(
N-K\right) \right] $, we get a vanishing probability of error. In
particular, under the assumption that $M\geq \Omega \left( K\log \frac{N}{K}%
\right) $, if $T\gg \frac{\log N}{\log \log N}$, then $\frac{\log \left[ K\left(
N-K\right) \right] }{\log \left[ K\log \frac{N}{K}\right] }\leq \frac{\log N}{%
\log \log N}$ implies that $K\left( N-K\right) \,\ll O\left( \left( \frac{K\log
\frac{N}{K}}{\sigma ^{2}}\right) ^{\kappa T}\right) =O\left( \frac{K\left(
N-K\right) }{\left( \frac{M}{\sigma ^{2}}\right) ^{\kappa T}}\right) $ for
suitably selected constants. $\blacksquare $

\bigskip We now consider several special cases and explore the implications
of the sufficient conditions. The discussions are heuristic in nature and
their validity requires further checking.

If we set $T=1$, then we need $M$ to be much greater than $N$ to guarantee a
vanishing probability $P_{\mathrm{err}}$. This restriction suggests that
even if we have more observations than the original signal length $N$, in
which case we can obtain the original sparse signal by solving a least
squares problem, we still might not be able to get the correct support
because of the noise, as long as $M$ is not sufficiently large compared to $N$. We
discussed in the introduction that for many applications, the support of a
signal has significant physical implications and its correct recovery is of
crucial importance. Therefore, without multiple temporal samples and with moderate noise, the scheme
proposed by Compressive Sensing is questionable as far as support recovery
is concerned. Worse, if we set the variance for the elements in $\boldsymbol{%
A}$ to be $1/M$ as in Compressive Sensing, which is equivalent to
replacing $\sigma ^{2}$ with $M\sigma ^{2}$, even increasing the
number of temporal samples will not improve the probability of
error significantly unless the noise variance is very small.
Hence, using support recovery as a criterion, one cannot expect
the Compressive Sensing scheme to work very well in the low SNR
case. This conclusion is not a surprise, since we reduce the
number of samples to achieve compression.

Another special case is when $K=1$. In this case, the sufficient condition
becomes $M\geq \log N$ and $\kappa T\log \frac{M}{\sigma ^{2}}\gg \log N.$ Now
the number of total samples should satisfy $MT\gg \frac{\left( \log N\right)
^{2}}{\log \log N}$ while the necessary condition states that $MT=\Omega
\left( \log N\right) .$ The smallest gap between the necessary condition and
sufficient condition is achieved when $K=1$.

From a denoising perspective, Fletcher \emph{et.al.} \cite{fletcher2006denoising} upper bounded and approximated the probability of error
 for support recovery averaged over the Gaussian ensemble. The bound and its approximation are applicable only to the special case with $K=1$ and involve complex integrals. The authors obtained interesting SNR threshold as a function of $M,N$ and $K$ through the analytical bound. Note that our bounds are valid for general $K$ and have a simple form. Besides, most of our derivation is conditioned on a realization of the Gaussian measurement ensemble. The conditioning makes more sense than averaging since in practice we usually make observations with fixed sensing matrix and varying signals and noise.

The result of Theorem \ref{thm_suff_Gaussian} also exhibits several interesting properties in the general case.
Compared with the necessary condition \eqref{necessary_prob_1} and %
\eqref{necessary_prob_2}, the asymmetry in the sufficient condition is even
more desirable in most cases because of the asymmetric cost associated with
sensors and temporal samples. Once the threshold $K\log \frac{N}{K}$ of $M$
is exceeded, we can achieve a desired probability of error by taking more
temporal samples. If we were concerned only with total the number of samples, we would minimize $MT$ subject to the constraints %
\eqref{sufficient_M} and \eqref{sufficient_TlogM} to achieve a given level
of probability of error. However, in applications for which timing is
important, one has to increase sensors to reduce $P_{\mathrm{err}}$ to a
certain limit.

The sufficient condition \eqref{sufficient_M}, \eqref{sufficient_TlogM}, and %
\eqref{sufficient_T} is separable in the following sense. We observe from
the proof that the requirement $M=\Omega \left( K\log \frac{N}{K}\right) $ is
used only to guarantee that the randomly generated measurement matrix is a
good one in the sense that its incoherence $\bar{\lambda}$ is sufficiently
large, as in the case of Compressive Sensing. It is in Lemma \ref{tailprob}
that we use the Gaussian ensemble assumption. If another deterministic
construction procedure (for attempts in this direction, see \cite%
{Devore2007Deterministic}) or random distribution give measurement matrix
with better incoherence $\bar{\lambda}$, it would be possible to reduce the
orders for both $M$ and $T$.

\section{Conclusions}
\label{sec:conclusion}
In this paper, we formulated the support recovery problems for
jointly sparse signals as binary and multiple-hypothesis testings.
Adopting the probability of error as the performance criterion,
the optimal decision rules are given by the likelihood ratio test
and the maximum \textit{a posteriori} probability estimator. The
latter reduces to the maximum likelihood estimator when equal
prior probabilities are assigned to the supports. We then employed
the Chernoff bound and Fano's inequality to derive bounds on the
probability of error. We discussed the implications of these
bounds at the end of Section \ref{sec:multipleupper},
Section \ref{sec:noiseeffectupper}, Section
\ref{sec:lower}, Section \ref{sec:gaussian_nece}, and
Section \ref{sec:gaussian_suff}, in particular when they are
applied to the DOA estimation problem considered in
\cite{Willsky2005Source} and Compressive Sensing with a Gaussian
measurement ensemble. We derived sufficient and necessary
conditions for Compressive Sensing using Gaussian measurement matrices to achieve a vanishing probability of error in both the
mean and large probability senses. These conditions show the
necessity of considering multiple temporal samples. The symmetric
and asymmetric roles played by the spatial and temporal samples
and their implications in system design were discussed. For
Compressive Sensing, we demonstrated that it is impossible to
obtain accurate signal support with only one temporal sample if
the variance for the Gaussian measurement matrix scales with
$1/M$ and there is considerable noise.

This research on support recovery for jointly sparse signals is far from
complete. Several questions remain to be answered. First, we notice an
obvious gap between the necessary and sufficient conditions even in the
simplest case with $K=1$. Better techniques need to be introduced to refine
the results. Second, as in the case for RIP, computation of the quantity $%
\bar{\lambda}$ for an arbitrary measurement matrix is extremely difficult.
Although we derive large derivation bounds on $\bar{\lambda}$ and compute
the expected value for $\bar{\lambda}_{S_{i},S_{j}}$ for the Gaussian
ensemble, its behaviors in both the general and Gaussian cases require
further study. Its relationship with RIP also needs to be clarified.
Finally, our lower bound derived from Fano's inequality identifies only the
effect of the total gain. The effect of the measurement matrix's
incoherence is elusive. The answers to these questions will enhance our
understanding of the measurement mechanism \eqref{model_scalar}.


%

\appendices

\section{Proof of Proposition \protect\ref{eig_count}}

In this proof, we focus on the case for which both $k_{0}\neq 0$ and $%
k_{1}\neq 0$. Other cases have similar and simpler proofs. The eigenvalues
of $H$ satisfy $|\lambda \eye_{M}-H|=0$, which is equivalent to $|\lambda
\Sigma _{1}-\Sigma _{0}|=0$. The substitution $\lambda =\mu + 1$ defines
\begin{equation*}
g\left( \mu \right) =\left\vert \left( \mu +1\right) \Sigma _{1}-\Sigma
_{0}\right\vert =\left\vert \mu \Sigma _{1}-\left( \Sigma _{0}-\Sigma
_{1}\right) \right\vert .
\end{equation*}%
The following algebraic manipulation
\begin{eqnarray*}
G &\triangleq &\Sigma _{0}-\Sigma _{1} \\
&=&A_{S_{0}}A_{S_{0}}^{\dagger }-A_{S_{1}}A_{S_{1}}^{\dagger } \\
&=&\left[ A_{S_{0}\cap S_{1}}A_{S_{0}\cap S_{1}}^{\dagger
}+A_{S_{0}\backslash S_{1}}A_{S_{0}\backslash S_{1}}^{\dagger }\right]  \\
&&\ \ \ -\left[ A_{S_{0}\cap S_{1}}A_{S_{0}\cap S_{1}}^{\dagger
}+A_{S_{1}\backslash S_{0}}A_{S_{1}\backslash S_{0}}^{\dagger }\right]  \\
&=&A_{S_{0}\backslash S_{1}}A_{S_{0}\backslash S_{1}}^{\dagger
}-A_{S_{1}\backslash S_{0}}A_{S_{1}\backslash S_{0}}^{\dagger }
\end{eqnarray*}%
leads to
\begin{eqnarray*}
g\left( \mu \right)  &=&\left\vert \mu \Sigma _{1}-G\right\vert  \\
&\mathbf{=}&\left\vert \Sigma _{1}\right\vert ^{\frac{1}{2}}\left\vert \mu %
\eye_{M}-\Sigma _{1}^{-\frac{1}{2}}G\Sigma _{1}^{-\frac{1}{2}\dagger
}\right\vert \left\vert \Sigma _{1}\right\vert ^{\frac{1}{2}}.
\end{eqnarray*}

Therefore, to prove the theorem, it suffices to show that $\Sigma _{1}^{-%
\frac{1}{2}}G\Sigma _{1}^{-\frac{1}{2}\dagger }$ has $k_{0}$ positive
eigenvalues, $k_{1}$ negative eigenvalues and $M-\left( k_{0}+k_{1}\right) $
zero eigenvalues or, put another way, $\Sigma _{1}^{-\frac{1}{2}}G\Sigma
_{1}^{-\frac{1}{2}\dagger }$ has inertia $\left( k_{0},k_{1},M-\left(
k_{0}+k_{1}\right) \right) $. The Sylvester's law of inertia (\cite{Horn1990Matrix}, Theorem 4.5.8, p. 223) states that the inertia of a
symmetric matrix is invariant under congruence transformations. Hence, we
need only to show that $G$ has inertia $\left( k_{0},k_{1},M-\left(
k_{0}+k_{1}\right) \right) $. Clearly $G=PQ^{\dagger }$ with $P=\left[
\begin{array}{cc}
A_{S_{0}\backslash S_{1}} & A_{S_{1}\backslash S_{0}}%
\end{array}%
\right] $ and $Q=\left[
\begin{array}{cc}
A_{S_{0}\backslash S_{1}} & -A_{S_{1}\backslash S_{0}}%
\end{array}%
\right] .$ To find the number of zero eigenvalues of $G$, we calculate the
rank of $G$. The non-degenerateness of measurement matrix $A$ implies that $%
\func{rank}\left( P\right) =\func{rank}\left( Q\right) =k_{0}+k_{1}.$
Therefore, $\ $from rank inequality (\cite{Horn1990Matrix}, Theorem 0.4.5,
p. 13),
\begin{eqnarray*}
&&\rank\left( P\right) +\rank\left( Q^{\dagger }\right) -\left(
k_{0}+k_{1}\right)  \\
&&\ \ \ \ \ \ \ \ \ \ \ \ \ \ \ \leq \rank\left( PQ^{\dag }\right)  \\
&&\ \ \ \ \ \ \ \ \ \ \ \ \ \ \ \ \ \ \ \ \ \leq \min \left\{ \func{rank}%
\left( P\right) ,\func{rank}\left( Q^{\dag }\right) \right\} ,
\end{eqnarray*}%
we conclude that $\func{rank}\left( G\right) =k_{0}+k_{1}.$

\bigskip

To count the number of negative eigenvalues of $G$, we use the Jocobi-Sturm
rule (\cite{Gohberg2005Indefinite}, Theorem A.1.4, p. 320), which states
that for an $M\times M$ symmetric matrix whose $j\mathrm{th}$ leading
principal minor has determinant $d_{j},j=1,\ldots ,M$, the number of
nonnegative eigenvalues is equal to the number of sign changes of sequence $\{1,d_{1},\ldots ,d_{M}\}$. We consider only the first $k_{0}+k_{1}$ leading
principal minors, since higher order minors have determinant $0$.

Suppose $I=\{1,\ldots ,k_{0}+k_{1}\}$ is an index set. Without loss of
generality, we assume that $P^{I}$ is nonsingular. Applying $QL$
factorization (one variation of $QR$ factorization, see \cite%
{Golub1996Matrix}) to matrix $P^{I}$, we obtain $P^{I}=OL$, where $O$ is an
orthogonal matrix, $OO^{\dagger }=\eye_{k_{0}+k_{1}}$, and $L=\left(
l_{ij}\right) _{\left( k_{0}+k_{1}\right) \times \left( k_{0}+k_{1}\right) }$
is an lower triangular matrix. The diagonal entries of $L$ are nonzero since
$P^{I}$ is nonsingular. The partition of $L$ into
\begin{equation*}
L=\left[
\begin{array}{cc}
L_{1} & L_{2}%
\end{array}%
\right]
\end{equation*}%
with $L_{1}\in \mathbb{F}^{\left( k_{0}+k_{1}\right) \times k_{0}},L_{2}\in
\mathbb{F}^{\left( k_{0}+k_{1}\right) \times k_{1}}$, and $L_{2}=\left[
\begin{array}{c}
0 \\
L_{3}%
\end{array}%
\right] $ with $L_{3}\in \mathbb{F}^{k_{1}\times k_{1}}$ implies
\begin{equation*}
G_{I}^{I}=P^{I}(Q_{I})^{\dagger }=O\left[
\begin{array}{cc}
L_{1} & L_{2}%
\end{array}%
\right] \left[
\begin{array}{c}
L_{1}^{\dagger } \\
-L_{2}^{\dagger }%
\end{array}%
\right] O^{\dagger }.
\end{equation*}%
Again using the invariance property of inertia under congruence
transformation, we focus on the leading principal minors of $U\triangleq %
\left[
\begin{array}{cc}
L_{1} & L_{2}%
\end{array}%
\right] \left[
\begin{array}{c}
L_{1}^{\dagger } \\
-L_{2}^{\dagger }%
\end{array}%
\right] .$ Suppose $J=\{1,\ldots ,j\}$. For $1\leq j\leq k_{0}$, from the
lower triangularity of $L$, it is clear that
\begin{equation*}
\left\vert \left( U_{J}^{J}\right) \right\vert =\left\vert
(L_{1})_{J}^{J}\right\vert ^{2}=\prod_{i=1}^{j}\left\vert l_{ii}\right\vert
^{2}>0.
\end{equation*}%
For $k_{0}+1\leq j\leq k_{0}+k_{1}$, suppose $J_{0}=\{1,\ldots ,k_{0}\}$ and
$J_{1}=\{1,\ldots ,j-k_{0}\}$. We then have
\begin{eqnarray*}
\left\vert U_{J}^{J}\right\vert  &=&\left\vert \left( L_{1}\right)
_{J_{0}}^{J_{0}}\right\vert ^{2}\left\vert \left( L_{3}\right)
_{J_{1}}^{J_{1}}\right\vert \left\vert -\left[ (L_{3})_{J_{1}}^{J_{1}}\right]
^{\dagger }\right\vert  \\
&=&\left( -1\right) ^{j-k_{0}}\left\vert \left( L_{1}\right)
_{J_{0}}^{J_{0}}\right\vert ^{2}\left\vert \left( L_{3}\right)
_{J_{1}}^{J_{1}}\right\vert ^{2} \\
&=&\left( -1\right) ^{j-k_{0}}\prod_{i=1}^{j}\left\vert l_{ii}\right\vert
^{2}.
\end{eqnarray*}%
Therefore, the sequence $1,d_{1},d_{2},\cdots d_{k_{0}+k_{1}}$ has $k_{1}$
sign changes, which implies that $G_{I}^{I}$---hence $G$---has $k_{1}$
negative eigenvalues. Finally, we conclude that the theorem holds for $H$.

\section{Proof of Proposition \protect\ref{eig_bound}}

We first prove the first claim. From the proof of Proposition
\ref{eig_count},
it suffices to show that the sorted positive eigenvalues of $\Sigma _{1}^{-%
\frac{1}{2}}G\Sigma _{1}^{-\frac{1}{2}\dagger }$ are greater than those of $%
\frac{1}{\sigma ^{2}}R_{33}R_{33}^{\dagger }$, where $G=A_{S_{0}\backslash
S_{1}}A_{S_{0}\backslash S_{1}}^{\dagger }-A_{S_{1}\backslash
S_{0}}A_{S_{1}\backslash S_{0}}^{\dagger }$. Since cyclic permutation of a
matrix product does not change its eigenvalues, we restrict ourselves to $%
\Sigma _{1}^{-1}G$. Consider the $QR$ decomposition
\begin{eqnarray*}
&&%
\begin{bmatrix}
A_{S_{1}\backslash S_{0}} & A_{S_{1}S_{0}} & A_{S_{0}\backslash S_{1}}%
\end{bmatrix}%
=QR \\
&\triangleq &%
\begin{bmatrix}
Q_{1} & Q_{2} & Q_{3} & Q_{4}%
\end{bmatrix}%
\begin{bmatrix}
R_{11} & R_{12} & R_{13} \\
0 & R_{22} & R_{23} \\
0 & 0 & R_{33} \\
0 & 0 & 0%
\end{bmatrix}%
,
\end{eqnarray*}%
where $Q\in \mathbb{F}^{M\times M}$ is an orthogonal matrix with partitions $%
Q_{1}\in \mathbb{F}^{M\times k_{1}},Q_{2}\in \mathbb{F}^{M\times
k_{\mathrm{i}}},Q_{3}\in \mathbb{F}^{M\times k_{0}}$, $R\in \mathbb{F}^{M\times
\left( k_{1}+k_{\mathrm{i}}+k_{0}\right) }$ is an upper triangular matrix with
partitions $R_{11}\in \mathbb{F}^{k_{1}\times k_{1}},R_{22}\in \mathbb{F}%
^{k_{\mathrm{i}}\times k_{\mathrm{i}}},R_{33}\in \mathbb{F}^{k_{0}\times k_{0}}$, and other
submatrices have corresponding dimensions.

First, we note that%
\begin{eqnarray*}
&&Q^{\dagger }GQ \\
&=&%
\begin{bmatrix}
R_{13} & R_{11} \\
R_{23} & 0 \\
R_{33} & 0 \\
0 & 0%
\end{bmatrix}%
\begin{bmatrix}
R_{13}^{\dagger } & R_{23}^{\dagger } & R_{33}^{\dagger } & 0 \\
-R_{11}^{\dagger } & 0 & 0 & 0%
\end{bmatrix}
\\
&=&%
\begin{bmatrix}
\begin{bmatrix}
R_{13} & R_{11} \\
R_{23} & 0%
\end{bmatrix}%
\begin{bmatrix}
R_{13}^{\dagger } & R_{23}^{\dagger } \\
-R_{11}^{\dagger } & 0%
\end{bmatrix}
&
\begin{bmatrix}
R_{13}R_{33}^{\dagger } \\
R_{23}R_{33}^{\dagger }%
\end{bmatrix}
& 0 \\
\begin{bmatrix}
R_{33}R_{13}^{\dagger } & R_{33}R_{23}^{\dagger }%
\end{bmatrix}
& R_{33}R_{33}^{\dagger } & 0 \\
0 & 0 & 0%
\end{bmatrix}%
.
\end{eqnarray*}%
Therefore, the last $M-\left( k_{1}+k_{\mathrm{i}}+k_{0}\right) $ rows and columns of
$Q^\dagger G Q$---and hence of $(Q^\dagger \Sigma _{1}Q)^{-1}(Q^\dagger G Q)$---are zeros, which lead to the $%
M-\left( k_{1}+k_{\mathrm{i}}+k_{0}\right) $ zero eigenvalues of $\Sigma _{1}^{-\frac{%
1}{2}}G\Sigma _{1}^{-\frac{1}{2}\dagger }$. We then drop these rows and
columns in all matrices involved in subsequent analysis. In particular, the
submatrix of $Q^{\dagger }\Sigma _{1}Q=Q^{\dagger }\left( \sigma ^{2}\eye%
_{M}+A_{S_{1}}A_{S_{1}}^{\dagger }\right) Q$ without the last $M-\left(
k_{1}+k_{\mathrm{i}}+k_{0}\right) $ rows and columns is\bigskip
\begin{eqnarray*}
&&\sigma ^{2}\eye_{M}+%
\begin{bmatrix}
R_{11} & R_{12} \\
0 & R_{22} \\
0 & 0%
\end{bmatrix}%
\begin{bmatrix}
R_{11}^{\dagger } & 0 & 0 \\
R_{12}^{\dagger } & R_{22}^{\dagger } & 0%
\end{bmatrix}
\\
&=&%
\begin{bmatrix}
\sigma ^{2}\eye_{k_{1}+k_{\mathrm{i}}}+%
\begin{bmatrix}
R_{11} & R_{12} \\
0 & R_{22}%
\end{bmatrix}%
\begin{bmatrix}
R_{11}^{\dagger } & 0 \\
R_{12}^{\dagger } & R_{22}^{\dagger }%
\end{bmatrix}
& 0 \\
0 & \sigma ^{2}\eye_{k_{0}}%
\end{bmatrix}
\\
&\triangleq &%
\begin{bmatrix}
F & 0 \\
0 & \sigma ^{2}\eye_{k_{0}}%
\end{bmatrix}%
.
\end{eqnarray*}%
Define

\begin{eqnarray*}
&&%
\begin{bmatrix}
V & K^{\dagger } \\
K & R_{33}R_{33}^{\dagger }%
\end{bmatrix}
\\
&\triangleq &%
\begin{bmatrix}
\begin{bmatrix}
R_{13} & R_{11} \\
R_{23} & 0%
\end{bmatrix}%
\begin{bmatrix}
R_{13}^{\dagger } & R_{23}^{\dagger } \\
-R_{11}^{\dagger } & 0%
\end{bmatrix}
&
\begin{bmatrix}
R_{13}R_{33}^{\dagger } \\
R_{23}R_{33}^{\dagger }%
\end{bmatrix}
\\
\begin{bmatrix}
R_{33}R_{13}^{\dagger } & R_{33}R_{23}^{\dagger }%
\end{bmatrix}
& R_{33}R_{33}^{\dagger }%
\end{bmatrix}%
.
\end{eqnarray*}%
Due to the invariance of eigenvalues with respect to orthogonal
transformations and switching to the symmetrized version, we focus on
\begin{eqnarray*}
&&%
\begin{bmatrix}
F & 0 \\
0 & \sigma ^{2}\eye_{k_{0}}%
\end{bmatrix}%
^{-\frac{1}{2}}%
\begin{bmatrix}
V & K^{\dagger } \\
K & R_{33}R_{33}^{\dagger }%
\end{bmatrix}%
\begin{bmatrix}
F & 0 \\
0 & \sigma ^{2}\eye_{k_{0}}%
\end{bmatrix}%
^{-\frac{1}{2}\dagger } \\
&=&%
\begin{bmatrix}
F^{-\frac{1}{2}}VF^{-\frac{1}{2}\dagger } & F^{-\frac{1}{2}}K^{\dagger }%
\frac{1}{\sigma } \\
\frac{1}{\sigma }KF^{-\frac{1}{2}} & \frac{1}{\sigma ^{2}}%
R_{33}R_{33}^{\dagger }%
\end{bmatrix}.
\end{eqnarray*}

\bigskip
Next we argue that the sorted positive eigenvalues of $%
\begin{bmatrix}
F^{-\frac{1}{2}}VF^{-\frac{1}{2}\dagger } & F^{-\frac{1}{2}}K^{\dagger }%
\frac{1}{\sigma } \\
\frac{1}{\sigma }KF^{-\frac{1}{2}} & \frac{1}{\sigma ^{2}}%
R_{33}R_{33}^{\dagger }%
\end{bmatrix}%
$ are greater than the corresponding sorted eigenvalues of $\frac{1}{\sigma
^{2}}R_{33}R_{33}^{\dagger }$.

For any $\varepsilon >0$, we define a matrix $M_{\varepsilon ,N}=\left[
\begin{array}{cc}
-N\eye_{k_{1}+k_{\mathrm{i}}} & 0 \\
0 & \frac{1}{\sigma ^{2}}R_{33}R_{33}^{\dagger }-\varepsilon \eye_{k_{0}}%
\end{array}%
\right] $. Then we have

\bigskip
\begin{eqnarray*}
&&%
\begin{bmatrix}
F^{-\frac{1}{2}}VF^{-\frac{1}{2}\dagger } & F^{-\frac{1}{2}}K^{\dagger }%
\frac{1}{\sigma } \\
\frac{1}{\sigma }KF^{-\frac{1}{2}} & \frac{1}{\sigma ^{2}}%
R_{33}R_{33}^{\dagger }%
\end{bmatrix}%
-M_{\varepsilon ,N} \\
&=&%
\begin{bmatrix}
F^{-\frac{1}{2}}VF^{-\frac{1}{2}\dagger }+N\eye_{k_{1}+k_{\mathrm{i}}} & F^{-\frac{1}{%
2}}K^{\dagger }\frac{1}{\sigma } \\
\frac{1}{\sigma }KF^{-\frac{1}{2}} & \varepsilon \eye_{k_{0}}%
\end{bmatrix}.%
\end{eqnarray*}%
Note that $%
\begin{bmatrix}
F^{-\frac{1}{2}}VF^{-\frac{1}{2}\dagger }+N\eye_{k_{1}+k_{\mathrm{i}}} & F^{-\frac{1}{%
2}}K^{\dagger }\frac{1}{\sigma } \\
\frac{1}{\sigma }KF^{-\frac{1}{2}} & \varepsilon \eye_{k_{0}}%
\end{bmatrix}%
$ is congruent to

\begin{equation*}
\begin{bmatrix}
F^{-\frac{1}{2}}VF^{-\frac{1}{2}\dagger }+N\eye_{k_{1}+k_{\mathrm{i}}}-\frac{1}{%
\varepsilon \sigma ^{2}}F^{-\frac{1}{2}}K^{\dagger }KF^{-\frac{1}{2}} & 0 \\
0 & \varepsilon \eye_{k_{0}}%
\end{bmatrix}%
.
\end{equation*}%
Clearly $F^{-\frac{1}{2}}VF^{-\frac{1}{2}\dagger }+N\eye_{k_{1}+k_{\mathrm{i}}}-\frac{%
1}{\varepsilon \sigma ^{2}}F^{-\frac{1}{2}}K^{\dagger }KF^{-\frac{1}{2}}$ is
positive definite when $N$ is sufficiently large. Hence, when $N$ is large
enough, we obtain
\begin{equation*}
\begin{bmatrix}
F^{-\frac{1}{2}}VF^{-\frac{1}{2}\dagger } & F^{-\frac{1}{2}}K^{\dagger }%
\frac{1}{\sigma } \\
\frac{1}{\sigma }KF^{-\frac{1}{2}} & \frac{1}{\sigma ^{2}}%
R_{33}R_{33}^{\dagger }%
\end{bmatrix}%
\succ M_{\varepsilon ,N}.
\end{equation*}%
Using Corollary 4.3.3 of \cite{Horn1990Matrix}, we conclude that the
eigenvalues of $%
\begin{bmatrix}
F^{-\frac{1}{2}}VF^{-\frac{1}{2}\dagger } & F^{-\frac{1}{2}}K^{\dagger }%
\frac{1}{\sigma } \\
\frac{1}{\sigma }KF^{-\frac{1}{2}} & \frac{1}{\sigma ^{2}}%
R_{33}R_{33}^{\dagger }%
\end{bmatrix}%
$ are greater than those of $M_{\varepsilon ,N}$ if sorted. From Proposition \ref%
{eig_count}, we know that $%
\begin{bmatrix}
F^{-\frac{1}{2}}VF^{-\frac{1}{2}\dagger }+N\eye_{k_{1}+k_{\mathrm{i}}} & F^{-\frac{1}{%
2}}K^{\dagger }\frac{1}{\sigma } \\
\frac{1}{\sigma }KF^{-\frac{1}{2}} & \varepsilon \eye_{k_{0}}%
\end{bmatrix}%
$ has exactly $k_{0}$ positive eigenvalues, which are the only eigenvalues
that could be greater than $\lambda \left( \frac{1}{\sigma ^{2}}%
R_{33}R_{33}^{\dagger }\right) -\varepsilon $. Since $\varepsilon $ is
arbitrary, we finally conclude that the positive eigenvalues of $\Sigma
_{1}^{-1}G$ are greater than those of $\frac{1}{\sigma ^{2}}%
R_{33}R_{33}^{\dagger }$ if sorted in the same way.

\bigskip For the second claim, we need some notations and properties of symmetric and Hermitian matrices. For any pair of
symmetric (or Hermitian) matrices $P$ and $Q$, $P\prec Q$ means that $Q-P$
is positive definite and $P\preceq Q$ means $Q-P$ is nonnegative definite.
Note that if $P$ and $Q$ are positive definite, then from Corollary 7.7.4 of
\cite{Horn1990Matrix} $P\preceq Q$ if and only if $Q^{-1}\preceq P^{-1}$; if $%
P\preceq Q$ then the eigenvalues of $P$ and $Q$ satisfy $\lambda _{k}\left(
P\right) \leq \lambda _{k}\left( Q\right) $, where $\lambda_k(P)$ denotes the $k$th largest eigenvalue of $P$; furthermore, $A\preceq B$ implies that $PAP^\dagger \preceq PBP^\dagger$ for any $P$, square or rectangular. Therefore, $\sigma ^{2}\eye%
_{M}+A_{S_{0}S_{1}}A_{S_{0}S_{1}}^{\dagger }\preceq \sigma ^{2}\eye%
_{M}+A_{S_{1}}A_{S_{1}}^{\dagger }=\Sigma _{1}\ $yields
\begin{equation*}
\Sigma _{0}^{1/2}\Sigma _{1}^{-1}\Sigma _{0}^{1/2}\preceq \Sigma
_{0}^{1/2}\left( \sigma ^{2}\eye_{M}+A_{S_{0}S_{1}}A_{S_{0}S_{1}}^{\dagger
}\right) ^{-1}\Sigma _{0}^{1/2}.
\end{equation*}

Recall that from the definition of
eigenvalues, the non-zero eigenvalues of $AB$ and $BA$ are the same for any
matrices $A$ and $B$. Since we are interested only in the eigenvalues, a cyclic permutation in the
matrix product on the previous inequality's right-hand side gives us
\begin{eqnarray*}
&&\left( \sigma ^{2}\eye_{M}+A_{S_{0}S_{1}}A_{S_{0}S_{1}}^{\dagger }\right)
^{-\frac{1}{2}}\Sigma _{0}\left( \sigma ^{2}\eye%
_{M}+A_{S_{0}S_{1}}A_{S_{0}S_{1}}^{\dagger }\right) ^{-\frac{1}{2}} \\
&=&\eye_{M}+\left( \sigma ^{2}\eye_{M}+A_{S_{0}S_{1}}A_{S_{0}S_{1}}^{\dagger
}\right) ^{-\frac{1}{2}}A_{S_{0}\backslash S_{1}} \\
&&\ \ \ \ \ \ \ \ \ \ \ \ \ \ \ \times A_{S_{0}\backslash S_{1}}^{\dagger
}\left( \sigma ^{2}\eye_{M}+A_{S_{0}S_{1}}A_{S_{0}S_{1}}^{\dagger }\right)
^{-\frac{1}{2}} \\
&\triangleq & \eye_{M} + Q^{-\frac{1}{2}}A_{S_{0}\backslash S_{1}}A_{S_{0}\backslash S_{1}}^{\dagger }Q^{-\frac{1}{2}}\\
&\triangleq &\eye_{M}+P.
\end{eqnarray*}

Until now we have shown that the sorted eigenvalues of $H$ are less than the
corresponding ones of $\eye_{M}+P$. The non-zero eigenvalues of $Q^{-\frac{1}{2}}A_{S_{0}\backslash S_{1}}A_{S_{0}\backslash S_{1}}^{\dagger }Q^{-\frac{1}{2}}$ is
the same as the non-zero eigenvalues of $A_{S_{0}\backslash S_{1}}^{\dagger }Q^{-1}A_{S_{0}\backslash S_{1}} \preceq \frac{1}{\sigma ^{2}} A_{S_{0}\backslash S_{1}}^{\dagger }A_{S_{0}\backslash S_{1}}$. Using the same fact again, we conclude that the non-zero eigenvalues of $\frac{1}{\sigma ^{2}} A_{S_{0}\backslash S_{1}}^{\dagger }A_{S_{0}\backslash S_{1}}$ is the same as the non-zero eigenvalues of $\frac{1}{\sigma ^{2}} A_{S_{0}\backslash S_{1}}A_{S_{0}\backslash S_{1}}^{\dagger }$. Therefore, we obtain that
\begin{eqnarray*}
  \lambda_k(\Sigma _{0}^{1/2}\Sigma _{1}^{-1}\Sigma _{0}^{1/2}) \leq \lambda_k(\eye_M + P) \leq \lambda_k\big(\eye_M + \frac{1}{\sigma ^{2}} A_{S_{0}\backslash S_{1}}A_{S_{0}\backslash S_{1}}^{\dagger }\big).
\end{eqnarray*}

In particular, the eigenvalues of
$H$ that are greater than $1$ are upper bounded by the corresponding ones of\ $\eye_M + \frac{1}{\sigma ^{2}} {A_{S_{0}\backslash S_{1}}A_{S_{0}\backslash S_{1}}^{\dagger }}$ if they are both sorted ascendantly. Hence, we get that the eigenvalues of $H$ that are
greater than $1$ are less than those of $\eye_{k_{0}}+\frac{1}{\sigma ^{2}}A_{S_{0}\backslash
S_{1}}^{\dagger }A_{S_{0}\backslash S_{1}}$.

Therefore, the conclusion of the second part of the theorem holds. We
comment here that usually it is not true that $H\preceq \eye_{M}+\frac{1}{%
\sigma ^{2}}A_{S_{0}\backslash S_{1}}A_{S_{0}\backslash S_{1}}^{\dagger }.$
Only the inequality on eigenvalues holds. $\ \ \ \ \ \ \ \blacksquare $

\section{Proof of Lemma \protect\ref{tailprob}}

For arbitrary fixed supports $S_{i},S_{j}$, we have

\begin{eqnarray*}
\bar{\lambda}_{S_{i},S_{j}} &\geq &1+\frac{1}{2\kappa \sigma ^{2}}\left(
\prod_{l=1}^{k_{\mathrm{d}}}2\kappa r_{ll}^{2}\right) ^{1/k_{\mathrm{d}}} \\
&\geq &\frac{1}{2\kappa \sigma ^{2}}\min_{1\leq l\leq k_{\mathrm{d}}}q_{l},
\end{eqnarray*}%
where $2\kappa r_{ll}^{2}\sim \chi _{2\kappa (M-K-l+1)}^{2}$ can
be written as a sum of $2\kappa (M-K-l+1)$ independent squared standard
Gaussian random variables and $q_{l}\sim \chi _{2\kappa
(M-2K)}^{2}$ is obtained by dropping $K-l+1$ of them. Therefore, using the
union bound we obtain
\begin{eqnarray*}
&&\Pr \left\{ \bar{\lambda}_{S_{i},S_{j}}\leq \gamma \right\}  \\
&\leq &\Pr \left\{ \frac{1}{2\kappa \sigma ^{2}}\min_{1\leq l\leq
k_{\mathrm{d}}}q_{l}\leq \gamma \right\}  \\
&\leq &\Pr \left\{ \bigcup_{1\leq l\leq k_{\mathrm{d}}}\left[ q_{l}\leq
2\kappa \sigma ^{2}\gamma \right] \right\}  \\
&\leq &k_{\mathrm{d}}\Pr \left\{ q_{l}\leq 2\kappa \sigma ^{2}\gamma \right\} .
\end{eqnarray*}%
Since $\gamma =\frac{1}{3}\frac{M-2K}{\sigma ^{2}}$ implies that $2\kappa
\sigma ^{2}\gamma =\frac{2\kappa }{3}\left( M-2K\right) <2\kappa (M-2K)-2$,
the mode of $\chi _{\kappa \left( M-2K\right) }^{2}$, when $M-2K$ is
sufficiently large, we have
\begin{eqnarray*}
&&\Pr \left\{ q_{l}\leq 2\kappa \sigma ^{2}\gamma \right\}  \\
&=&\int_{0}^{2\kappa \sigma ^{2}\gamma }\frac{\left( 1/2\right) ^{\kappa
(M-2K)}}{\Gamma \left( \kappa (M-2K)\right) }x^{\kappa (M-2K)-1}e^{-x/2}dx \\
&\leq &\frac{\left[ \kappa \sigma ^{2}\gamma \right] ^{\kappa (M-2K)}}{%
\Gamma \left( \kappa (M-2K)\right) }e^{-\kappa \sigma ^{2}\gamma }.
\end{eqnarray*}%
The inequality $\log \Gamma \left( z\right) \geq \left( z-\frac{1}{2}\right)
\log z-z$ says that when $M-2K$ is large enough,
\begin{eqnarray*}
&&\Pr \left\{ q_{l}\leq 2\kappa \sigma ^{2}\gamma \right\}  \\
&\leq &\exp {\Huge \{}\kappa (M-2K)\log \left( \kappa \sigma ^{2}\gamma
\right) -\kappa \sigma ^{2}\gamma  \\
&&\ \ \ \ \ \ \ \ \ \ -\left[ \kappa (M-2K)-\frac{1}{2}\right] \log \left[
\kappa (M-2K)\right]  \\
&&\ \ \ \ \ \ \ \ \ \ \ \ \ \ \ \ \ \ +\kappa (M-2K){\Huge \}} \\
&\leq &\exp \left\{ -c(M-2K)\right\} ,
\end{eqnarray*}%
where $c<\kappa \left( \log 3-1\right) $. Therefore, we have
\begin{equation*}
\Pr \left\{ \bar{\lambda}_{S_{i},S_{j}}\leq \gamma \right\} \leq K\exp
\left\{ -c(M-2K)\right\} .\ \ \ \ \ \ \blacksquare
\end{equation*}

\section*{Acknowledgment}
The authors thank the anonymous referees  for their careful and helpful comments.


\begin{IEEEbiographynophoto}{Gongguo Tang}
earned his B.Sc. degree in Mathematics from the Shandong
University, China in 2003, and the M.Sc. degree in System Science
from Chinese Academy of Sciences, China, in 2006.

Currently, he is a Ph.D. candidate with the Department of Electrical
and Systems Engineering, Washington University, under the guidance
of Dr. Arye Nehorai. His research interests are in the area of
Compressive Sensing, statistical signal processing, detection and
estimation, and their applications.
\end{IEEEbiographynophoto}

\begin{IEEEbiographynophoto}{Arye Nehorai}
(S'80-M'83-SM'90-F'94) earned his B.Sc. and M.Sc. degrees in
electrical engineering from the Technion--Israel Institute of
Technology, Haifa, Israel, and the Ph.D. degree in electrical
engineering from Stanford University, Stanford, CA.

From 1985 to 1995, he was a Faculty Member with the Department of
Electrical Engineering at Yale University. In 1995, he became a
Full Professor in the Department of Electrical Engineering and
Computer Science at The University of Illinois at Chicago (UIC).
From 2000 to 2001, he was Chair of the Electrical and Computer
Engineering (ECE) Division, which then became a new department. In
2001, he was named University Scholar of the University of
Illinois. In 2006, he became Chairman of the Department of
Electrical and Systems Engineering at Washington University in St.
Louis. He is the inaugural holder of the Eugene and Martha Lohman
Professorship and the Director of the Center for Sensor Signal and
Information Processing (CSSIP) at WUSTL since 2006.

Dr. Nehorai was Editor-in-Chief of the IEEE TRANSACTIONS ON SIGNAL
PROCESSING from 2000 to 2002. From 2003 to 2005, he was Vice
President (Publications) of the IEEE Signal Processing Society
(SPS), Chair of the Publications Board, member of the Board of
Governors, and member of the Executive Committee of this Society.
From 2003 to 2006, he was the founding editor of the special
columns on Leadership Reflections in the IEEE Signal Processing
Magazine. He was co-recipient of the IEEE SPS 1989 Senior Award
for Best Paper with P. Stoica, coauthor of the 2003 Young Author
Best Paper Award, and co-recipient of the 2004 Magazine Paper
Award with A. Dogandzic. He was elected Distinguished Lecturer of
the IEEE SPS for the term 2004 to 2005 and received the 2006 IEEE
SPS Technical Achievement Award. He is the Principal Investigator
of the new multidisciplinary university research initiative (MURI)
project entitled Adaptive Waveform Diversity for Full Spectral
Dominance. He has been a Fellow of the Royal Statistical Society
since 1996.
\end{IEEEbiographynophoto}





\end{document}